\documentclass[aps,pre,twocolumn,nobibnotes,superscriptaddress,longbibliography]{revtex4-2}
\usepackage{hyperref}
\usepackage{multirow}
\usepackage{float}
\usepackage{colortbl}
\usepackage{xcolor}
\usepackage{graphicx}
\usepackage{dcolumn}
\usepackage{bm}
\usepackage{booktabs}
\usepackage{makecell}
\usepackage{amsmath}
\usepackage{cases}
\usepackage{float}

\begin{document}
	\preprint{APS/123-QED}

\title{Role of chloride concentration in modulating seizure transitions in excitatory and inhibitory networks} 

\author{Qianchen Gong}
\altaffiliation{School of physics and Electronic Engineering, Jiangsu University, Zhenjiang, Jiangsu 212013, China}  
\author{Yingpeng Liu}
\altaffiliation{School of Mathematical Sciences, Jiangsu University, Zhenjiang 212013, Jiangsu, China} 
\author{Yan Zhang}
\altaffiliation{Teaching Department, Jiangsu University Jingjiang College, Zhenjiang 212028, Jiangsu, China } 
\author{Muhua Zheng}
\altaffiliation{School of physics and Electronic Engineering, Jiangsu University, Zhenjiang, Jiangsu 212013, China} 
\author{Kesheng Xu}
\altaffiliation{School of physics and Electronic Engineering, Jiangsu University, Zhenjiang, Jiangsu 212013, China}  
\email{ksxu@ujs.edu.cn}

\date{\today}

\begin{abstract}
Experimental evidence indicates that intracellular chloride concentration regulates the excitation–inhibition (EI) balance, yet the mechanisms by which activity-dependent chloride dynamics drive seizure evolution and stage transitions remain unclear. We present a conductance-based neuronal network in which EI balance emerges from chloride homeostasis via channel-mediated influx and transporter-mediated extrusion. We show that the fraction of inhibitory synaptic conductance contributing to channel-mediated influx acts as a control parameter that organizes seizure dynamics into distinct stages--pre-ictal, ictal-tonic, and ictal-clonic--distinguished by characteristic amplitude and frequency signatures. Decreasing this fraction shortens ictal activity and suppresses seizure initiation, whereas high fraction promotes the emergence of ictal-tonic and ictal-clonic stages and spiral-wave dynamics, rendering seizure dynamics largely insensitive to inhibition. At intermediate values, seizures bypass the ictal--tonic stage and emerge directly as the ictal--clonic stage. Moreover, joint variation of fractions with synaptic strengths reveals that recurrent excitation expands the tonic--clonic seizure, while recurrent inhibition prolongs pre-ictal states and suppresses ictal-clonic activity.

\end{abstract}

\maketitle

\section{Introduction}
Epilepsy is a neurological disorder characterized by seizures  typically originating in cortical or subcortical gray matter \cite{jirsa2014nature}. A hallmark of epilepsy is hypersynchronization,whereby large populations of neurons discharge excessively and simultaneously, producing global, high-amplitude patterns in electroencephalographic (EEG) recordings \cite{karpov2021noise}.Gastaut and Broughton classified seizures into pre-ictal, ictal-tonic, and ictal-clonic phases \cite{theodore1994secondarily}.The pre-ictal phase features brief periodic spikes, followed by an ictal-tonic phase with low-voltage, fast-onset gamma activity lasting about 30 seconds, and an ictal-clonic phase characterized by repetitive discharges with tremors of increasing amplitude and progressively decreasing frequency \cite{koksal2024expansion,zhang2024multimodal}.

Extensive research has shown that epilepsy is driven by an imbalance between excitation and inhibition (EI) toward excitation (hyperexcitability) within cortical networks\cite{milligan2021epilepsy,sumadewi2023biomolecular, untiet2024astrocyte}. A major contributor to seizure onset is the dysregulation of EI balance resulting from the interplay between excitatory synaptic coupling and inhibitory synaptic transmission in the central nervous system\cite{salners2024simple,untiet2024astrocyte,zhu2024chaos}. In addition, other mechanisms, including inflammation\cite{rana2018role}, ionic dysregulation \cite{kostrzewa2023handbook}, aberrant neurogenesis \cite{staley2015molecular}, genetic factors\cite{klein2024new}, further highlight the multifactorial nature of epilepsy. Together, these factors converge to destabilize excitability and drive hypersynchronous activity.

The vast majority of experimental and modeling studies have evidenced that the biophysical underpinnings of $\mathrm{Cl^-}$ homeostasis are tightly and dynamically regulated by the gamma-aminobutyric acid-ergic (GABAergic) chloride channels\cite{van2023excitatory, raimondo2017neuronal} and cation–chloride cotransporters (CCCs) \cite{delpire2025chloride,pressey2023chloride}.The GABAergic transmission, working for inhibitory synaptic coupling, activates chloride-permeable GABA receptors and opens ligand-gated $\mathrm{Cl^-}$ channels, which are selectively permeable to induce chloride ion flux ($\mathrm{Cl^-}$)\cite{zaccariello2025inhibitory}and, to a lesser extent, bicarbonate ions ($\mathrm{HCO_{3}^{-}}$)\cite{liu2020role}. The CCCs transport chloride ions and ions of $\mathrm{Na^+}$ and $\mathrm{K^+}$ together across the membranes surrounding cells,thereby establishing the chloride driving force \cite{currin2020chloride}. For example, the potassium–chloride cotransporter 2 (KCC2) mediates chloride extrusion and the sodium–potassium–chloride cotransporter 1 (NKCC1) promotes $\mathrm{Cl^-}$ influx in neurons \cite{pressey2023chloride}. Through these two mechanisms, intracellular chloride concentration exerts bidirectional control over both inhibitory synaptic coupling and CCCs, thereby regulating the balance between excitation and inhibition in cortical circuits and serving as a key homeostatic process that constrains the spread of excitation.

To date, only a surprisingly small number of well-characterized, biophysically grounded neuronal mechanisms have been shown to account for the complex, large-scale spatiotemporal evolution of focal seizures. To address this challenge, Liou et al. \cite{liou2020model} proposed a parsimonious, mechanistically informed framework that captures generalizable seizure dynamics. Their model formulates a three stage model of focal seizure evolution -- onset, propagation, and termination -- driven by shifts in the EI balance through spatial summation and changes in the intracellular chloride concentration. Within this framework, a transient excitatory perturbation initiates a localized tonic firing region (ictal-tonic stage), which subsequently undergoes a spontaneous transition to a bursting attractor (ictal-clonic stage). As seizure termination approaches, the ictal wavefront dissipates,leading to a pre-termination regime characterized by weakened bursting, followed by an abrupt return to a quiescent state after the final burst propagates through the network (post-ictal stage). However, the processes by which this network generates and transitions between these complex stages are still under investigation.

In this work, intracellular chloride dynamics are modeled as the result of two counteracting mechanisms: chloride influx through ligand-gated ion channels and chloride clearance via neuronal transporters operating with first-order kinetics  as previous investigations\cite{liou2020model,currin2020chloride,deisz2011components,currin2022computational}.For ligand-gated ion channels, GABAergic inhibitory synapses are selectively permeable to both $\mathrm{Cl^-}$ and $\mathrm{HCO_3^-}$ where the total number of receptors is fixed, with a defined fraction of the total GABAergic current carried by $\mathrm{Cl^-}$\cite{dusterwald2018biophysical,deisz2011components,currin2022computational}. Accordingly, the total GABAergic ionic conductance is decomposed into $\mathrm{Cl^-}$ and $\mathrm{HCO_3^-}$ components, weighted by their respective fractional contributions. However, the role of the relative $\mathrm{Cl^-}$ and $\mathrm{HCO_3^-}$ conductance fractions in shaping chloride influx underlying synaptic inhibition and in controlling seizure stage transitions within a modeling framework remains insufficiently explored. In particular, the mechanistically informed framework proposed by Liou et al. \cite{liou2020model} considers only pure $\mathrm{Cl^-}$ conductance, neglecting the contribution of $\mathrm{HCO_3^-}$. To address this issue, we introduce a control parameter $W_{con}$, representing the fraction of the total GABAergic ionic conductance (equivalently, the fraction of the total GABAergic current carried by $\mathrm{Cl^-}$) that regulates intracellular chloride influx.

 By varying $W_{con}$ while keeping synaptic weights fixed, we observe that at high $W_{con}$, the network sequentially produces pre-ictal, ictal-tonic, and ictal-clonic stages. A slight reduction in $W_{con}$ narrows the ictal-tonic stage and shortens seizures during the ictal-clonic phase, though the characteristic stages remain visible. Further reduction causes focal seizures in the ictal-clonic state to persist as the network enters a spiral wave state. At intermediate values of $W_{con}$, the network bypasses the ictal-tonic stage, transitioning directly to ictal-clonic. At lower values of $W_{con}$, the neural network fails to trigger focal seizure activity. By varying both $W_{con}$ and synaptic weights, we find that increasing $J_{EE}$ expands the regions of both clonic and tonic–clonic seizures  due to hyperexcitability at different $W_{con}$ levels, suggesting that excessive excitatory drive broadens the range of seizure-like dynamics. In contrast, enhancing recurrent inhibition $J_{EI}$ enlarges the seizure offset region and shrinks the clonic region when $W_{con}$ is slightly reduced. However, at high $W_{con}$ values, seizure dynamics become largely insensitive to inhibitory strength, consistently exhibiting tonic–clonic patterns. These findings highlight the critical role of impaired chloride homeostasis (elevated $W_{con}$) and the excitation–inhibition imbalance ($J_{EE}$ or $J_{EI}$) in seizure onset, propagation, termination, and stage transitions, aligning with experimental evidence that cortical hyperexcitability destabilizes network dynamics \cite{cherubini2022dysregulation, tang2021role}.

\section{Two-dimensional neural network described by the conductance model }

\begin{figure}[ht]	
	\includegraphics[width=\linewidth]{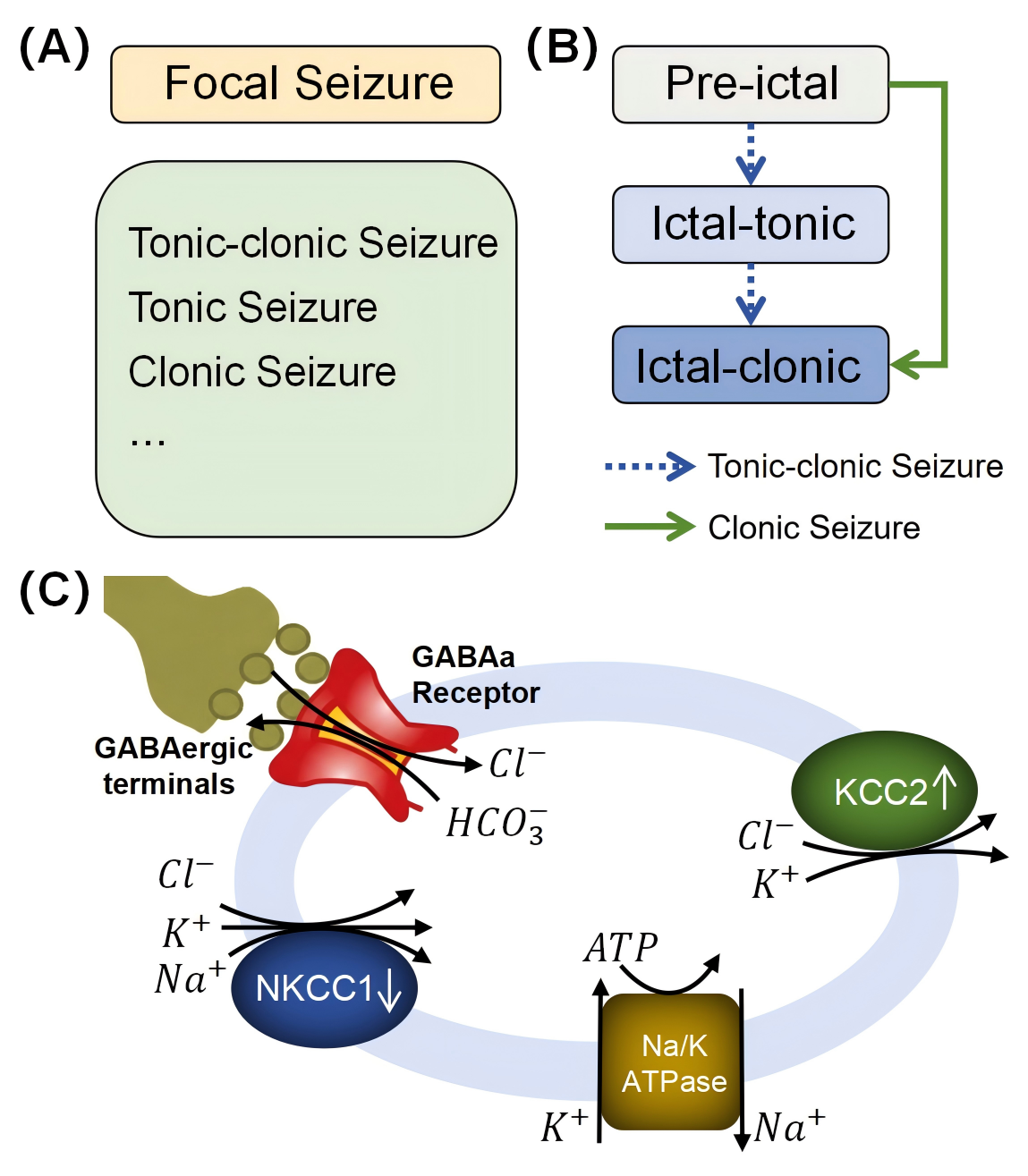}	  
	\caption{(A) Focal seizure classification. (B) Stages of focal seizure dynamics: pre-ictal, ictal-tonic, and ictal-clonic, observed in tonic-clonic and clonic seizures. (C) Chloride concentration regulation in GABAergic chloride channels and cation–chloride cotransporters.  Chloride transport is mediated by NKCC1 and KCC2, utilizing $\mathrm{Na^+}$ and $\mathrm{K^+}$ gradients by $\mathrm{Na^+}/\mathrm{K^+}$-ATPase. }
	\label{fig:clonic}
\end{figure}

Several seizure phenotypes have been characterized, including clonic, tonic, and tonic–clonic seizures (Fig.\hyperref[fig:clonic]{\ref{fig:clonic}A})\cite{thijs2021autonomic,fisher2017operational}. In this work, we develop an excitatory–inhibitory neural network model that reproduces the ictal-tonic and ictal-clonic stages, thereby giving rise to clonic and tonic–clonic seizures, as well as transitions between these seizure types (Fig.\hyperref[fig:clonic]{\ref{fig:clonic}B}). A key feature of the model is the introduction of a control parameter, $W_{con}$, which modulates intracellular chloride influx across different levels (Fig.\hyperref[fig:clonic]{\ref{fig:clonic}C}). This modulation enables the emergence of distinct seizure phenotypes, including clonic and tonic–clonic seizures.

\subsection{Conductance-based excitatory–inhibitory neural network}

\begin{figure}[htp]	
	\includegraphics[width=\linewidth]{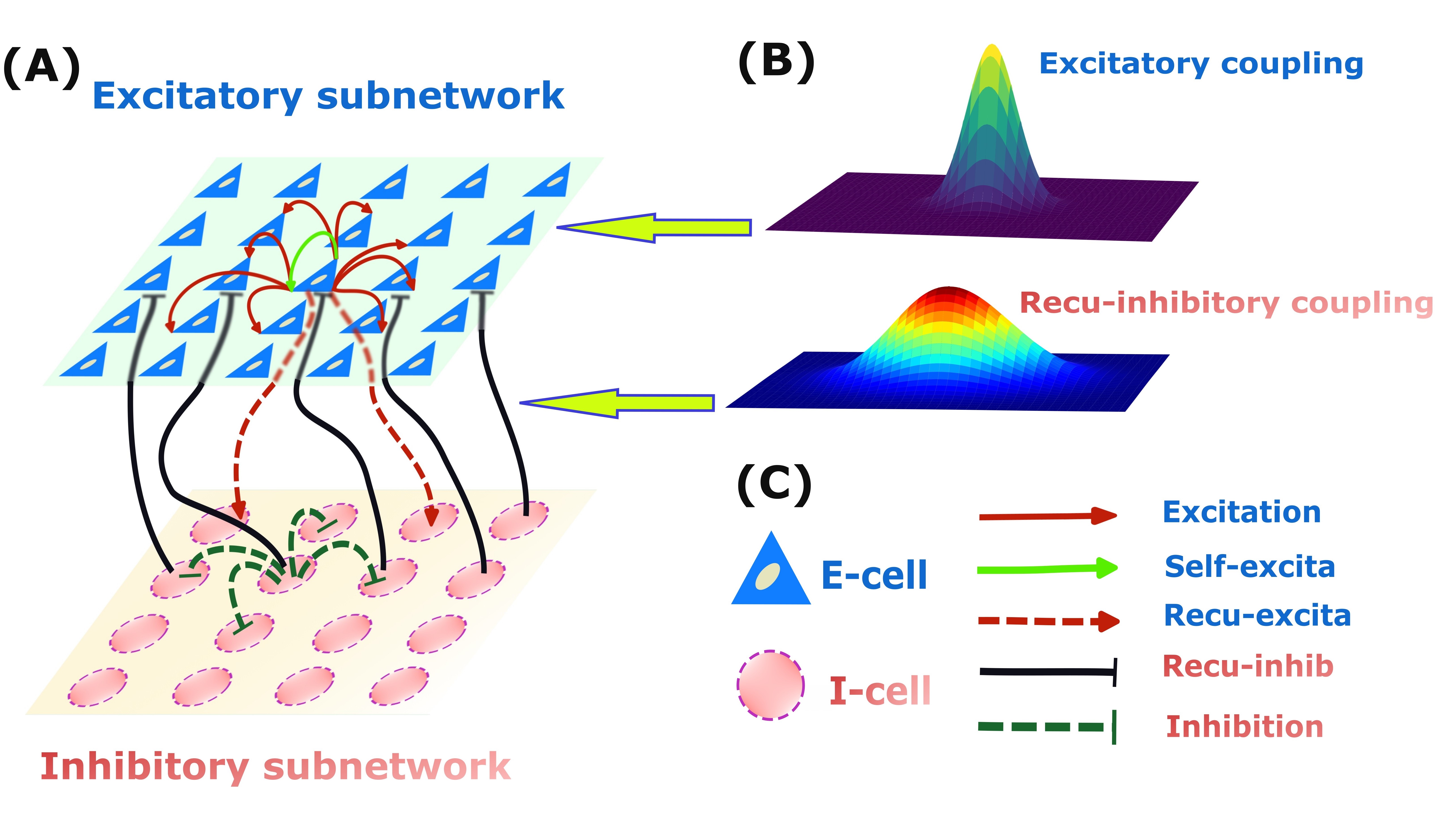}	  
	\caption{Two dimensional excitatory–inhibitory conductance-based network for focal seizure onset. (A) Two layers of excitatory and inhibitory networks. (B) Local excitatory and surround inhibitory synaptic coupling. (C) Neural types and synaptic connections. }
	\label{fig:EI}
\end{figure}

We consider a spatial neural field model with Mexican-hat connectivity (shown in Fig.\hyperref[fig:EI]{\ref{fig:EI}})\cite{amari1977dynamics,coombes2005waves,takiyama2016bayesian,morrison2021plastic}, which provides a biologically grounded description with local excitation and surround inhibition\cite{prince1967control}. This framework provides an effective description of mixed populations of interacting excitatory and inhibitory neurons with typical cortical connectivity, and is used to investigate the emergence of spatiotemporal seizure dynamics and their transitions. The homogeneous spatial excitatory neural field is defined within a partitioned square space of $N = 100 \times 100$ neurons, indexed by $i$ and $j$ [as shown in Fig.\hyperref[fig:EI]{\ref{fig:EI}A}], where $i$ and $j$ represent the neural indices from excitatory population\cite{liou2020model}. Each neuron receives recurrent synaptic currents from the spatial summation of excitatory and inhibitory subnetworks. Recurrent excitatory (red/green solid lines) and inhibitory (black solid lines) connections are computed via two-dimensional spatial Gaussian convolution(Fig.\hyperref[fig:EI]{\ref{fig:EI}B}). The time evolution of the excitatory neuron at position ($i,j$) is governed by a set of equations:

{\small{
\begin{subequations}
	\begin{align}   
		C\frac{dV_{ij}}{dt} & = -I_{L,ij}- I_{K,ij} -I_{\mathrm{ext}} -I_{sE,ij}- I_{Cl,ij} \label{Menpo} \\ 
		\tau_K\frac{ dg_{K_{ij}}}{dt} & =  -g_{K_{ij}} +\Delta_Kf(V_{ij}),  \label{thre_ada}\\
		\tau_\phi\frac{ d\phi_{ij}}{dt} & = (\phi_0 - \phi_{ij}) +\Delta_\phi f(V_{ij}),  \label{sAHP}\\
		\frac{d\text{Cl}_{in,ij}}{dt} & = \frac{I_{\text{Cl},ij}}{V_dF_r}W_{con} + \frac{\text{Cl}_{in,eq} -\text{Cl}_{in,ij}}{\tau_{\text{Cl}}}, \label{chlorideflow}
	\end{align}
	\label{EQ1}
\end{subequations}}}
Where,
	
\begin{subequations}
	\begin{align} 
	I_{L,ij} & =  g_L(V_{ij}-E_L)  \label{IL}\\
	I_{K,ij} & =  g_{K,{ij}}(V_{ij}-E_K)   \label{IK}\\
	I_{sE,ij} & =  g_{\mathrm{sE},{ij}}(V_{ij}-E_{\mathrm{sE}})   \label{IsE}\\
	I_{Cl,ij} & =   g_{\mathrm{Cl},{ij}}(V_{ij}- E_{\text{Cl},ij}) \label{ICL} \\
	E_{\text{Cl},ij} &= -26.7\text{log}\dfrac{\text{Cl}_{out}}{\text{Cl}_{in,ij}}  \label{ECL}  \\
	f(V_{ij}) & = \dfrac{f_{max}}{1+e^{[-(V_{ij}-\phi_{ij})/\beta]}}  \label{fvij}	 
\end{align}
\label{EQ2}
\end{subequations}	

The variables $V_{ij}$, $g_{K_{ij}}$, $\phi_{ij}$ and $\text{Cl}_{in,ij}$ represent the membrane potential, the slow afterhyperpolarization (sAHP) conductance, the firing threshold, and the intracellular chloride concentration, respectively. Eq.\hyperref[Menpo]{(\ref{Menpo})} describes a conductance-based neuronal model of neocortical pyramidal neurons, incorporating the leak current \cite{tripathy2014neuroelectro} as well as the activity-dependent emergence of chloride clearance\cite{deisz2011components}, buffering mechanisms\cite{marchetti2005modeling}, and sAHP\cite{alger1980epileptiform} observed during seizures. Thus, the terms $I_{L,ij}$, $I_{K,ij}$, $I_{ext}$, $I_{sE,ij}$, $I_{Cl,ij}$ in Eq.\hyperref[Menpo]{(\ref{Menpo})} represent the leak, sAHP, external input, excitatory synaptic, and GABAergic inhibitory synaptic currents, respectively [defined in Eqs.\hyperref[IL]{(\ref{IL})}–\hyperref[ICL]{(\ref{ICL})}]. Their associated conductances are expressed as $g_L$, $g_K$, $g_{\mathrm{sE}}$, and $g_{\mathrm{Cl}}$, with the corresponding reversal potentials denoted as $E_{L}$, $E_{K}$, $E_{\mathrm{sE}}$ and $E_{\mathrm{Cl}}$.  $C$ represents the cell capacitance. The default parameters in Eq.\hyperref[Menpo]{(\ref{Menpo})} based on phenomenological considerations\cite{tripathy2014neuroelectro,marchetti2005modeling,alger1980epileptiform} are: $C =100$ \text{pF}, $g_L = 4$ \text{nS}, $E_L = -58$ \text{mV},  $E_K = -90$ \text{mV}, $E_{\mathrm{sE}} = 0$ \text{mV}.
	
The evolution of the steady state of the sAHP conductance $g_{K}$  is linearly driven by the mean firing rate $f(V_{ij})$ through $\Delta_{K}$ according to first-order kinetics (see Eq.\hyperref[thre_ada]{(\ref{thre_ada})})\cite{alger1980epileptiform}, where the time constant $\tau_K$ is significantly larger than that of threshold adaptation  $\tau_{\phi}$. The mean firing rate $f(V_{ij})$ is defined in Eq.\hyperref[fvij]{(\ref{fvij})} as a sigmoid function of the difference between the membrane potential $V_{ij}$ and the firing threshold $\phi_{ij}$, where $f_{max}$ denotes the maximal population firing rate and $\beta$ controls the slope of the sigmoid function.  The firing threshold $\phi_{ij}$ also evolves toward a steady-state value that is linearly activated by the mean firing rate $f(V_{ij})$, with an amplitude determined by the coefficient $\Delta_\phi$, as described in Eq.\hyperref[sAHP]{(\ref{sAHP})}, where $\tau_{\phi}$ and $\phi_0$ denote the time constant and the baseline firing threshold of excitatory neurons, respectively \cite{deisz2011components}. We used $\phi_{0} = -45$ \text{mV}, $\tau_K  = 5$ \text{s}, $\tau_{\phi} = 100$ \text{ms}, $\Delta_{K} = 0.2$ nS/Hz,  $\Delta_{\phi} = 0.3$ mV/Hz, $\beta$ =2.5 \text{mV},$f_{\max} = 200$ \text{Hz} as previous studies\cite{tripathy2014neuroelectro,marchetti2005modeling,alger1980epileptiform}.

The reversal potential $E_{Cl}$ given in Eq.\hyperref[ECL]{(\ref{ECL})}, is determined by the Nernst equation and depends on the principal neuronal chloride gradient. The Eq.\hyperref[chlorideflow]{(\ref{chlorideflow})} describes GABA-mediated $Cl^-$ accumulation together with an exponential recovery toward the resting intracellular chloride level $\text{Cl}_{in,eq}$, governed by a clearance mechanism mediated by neuronal transporters with decay time constant $\tau_{Cl}$. In Eq.\hyperref[chlorideflow]{(\ref{chlorideflow})}, $I_{Cl,ij}$, $V_{d}$, $F_{r}$ denote the inhibitory chloride synaptic current [defined in Eq.\hyperref[ICL]{(\ref{ICL})}], the intracellular chloride distribution volume \cite{marchetti2005modeling, vladimirski2008episodic}, and Faraday’s constant, respectively, whereas $\tau_{Cl}$  and $\text{Cl}_{in,eq}$ represent the time constant of the chloride clearance mechanism and the equilibrium intracellular chloride concentration. The default parameters for intracellular chloride concentration are\cite{liou2020model}: $F_r= 9.649 \times 10^4 $ \text{C/mol}, $V_d = 0.24$ \text{pL},  $\tau_{\text{Cl}} = 5$ \text{s},  $\text{Cl}_{in,eq} = 6$ \text{mM}, $\text{Cl}_{out} = 110$ \text{mM}.

 More importantly, the total GABAergic ionic conductance is decomposed into $\mathrm{Cl^-}$ and $\mathrm{HCO_3^-}$ components, weighted by their respective fractional contributions shown in Fig.\hyperref[fig:clonic]{\ref{fig:clonic}C}. Thus, we introduce a control parameter $W_{con}$ in Eq.\hyperref[chlorideflow]{(\ref{chlorideflow})}, defined as the fraction of the total GABAergic ionic conductance (equivalently, the fraction of the total GABAergic current) carried by $\mathrm{Cl^-}$, which regulates intracellular chloride influx.
 
It is important to note that population dynamics of inhibitory interneurons in this study are simplified as outlined in Ref.\cite{liou2020model}. The  membrane potential dynamics of inhibitory neurons are not explicitly modeled; instead, they are assumed to respond instantaneously with a monotonic reaction to their synaptic inputs.

\subsection{Synaptic conductance of spatial summation and temporal summation}

The center–surround spatial coupling shown in Fig.\hyperref[fig:EI]{\ref{fig:EI}(B)} that represents the synaptic connectivity between neurons at locations ($i_{0},j_{0}$) and  $(i,j)$  is modeled by a Mexican-hat–shaped weight function, constructed as the difference of Gaussian functions \cite{singh2021mexican}.

In order to simulate the synaptic currents, the most general way for the recurrent excitatory synaptic conductance is to apply a spatial convolution of the normalized firing rate  $F_{ij}= f(V_{ij})/f_{max}$ with the  kernel $K^E \sim G(0, \sigma^2_{E})$, where $\sigma_E = 0.02$ is used in the main simulations (see Appendix A for additional values of $\sigma_E$). The result of this spatial convolution is then multiplied by the recurrent excitatory strength $J_{EE} = 100$, yielding:
\begin{equation}
	\hat{g}_{E,ij}(t) = J_{EE}\sum_{i_0=-k_E}^{k_E}\sum_{j_0=-k_E}^{k_E} F_{i-i_0,j-j_0} K^E_{i_0+k_E,j_0+k_E}
\end{equation}	

Recurrent inhibition $g^{\mathrm{synI}}_{ij}$ [in \hyperref[Menpo]{Eq.(\ref{Menpo})}] projected onto the excitatory population is decomposed into spatially localized and non-localized components.The localized component $ K^{\mathrm{Loc,I}}$ follows a Mexican-hat structure (Fig.\hyperref[fig:EI]{\ref{fig:EI}B}), representing surround inhibition.The existence of surround inhibition reflects the typical EI balance between short-range excitation and localized inhibition.The distance-independent recurrent inhibition pathway (referred to as global inhibition $K^{\mathrm{Glob,I}}$) is regulated by the parameter $\gamma$, inspired by some observations of large-scale inhibitory effects associated with focal seizure activity\cite{eissa2017cross,liou2018role}. Recurrent inhibition is thus defined in terms of both localized and global inhibitory components:
\begin{equation}
 K^I \sim K^{\mathrm{Loc,I}} + K^{\mathrm{Glob,I}} =  (1-\gamma)G(0,\sigma_I^2) + \gamma U . 
\end{equation} 
 
 Here $\gamma$ controls the relative contribution of each component,  $G(0, \sigma^2_{I})$ denotes a Gaussian kernel with $\sigma_{I} = 0.03$ (see Appendix A). Note that, recurrent inhibition extends over a larger area than recurrent excitation, with $\sigma_I > \sigma_E$ (Fig.\hyperref[fig:EI]{\ref{fig:EI}B}) due to local excitation and surround inhibition\cite{amari1977dynamics,coombes2005waves,takiyama2016bayesian,morrison2021plastic}. $U = \sum_{i=0}^{n}\sum_{j=0}^{n} F_{ij} / n^2$ represents a uniform distribution across the neural field. Similarly, the result of this spatial convolution for inhibitory integration is multiplied by the localized ($J_{EI}$) and non-localized components ($J_{In}$) of recurrent inhibitory strength. This yields:
 		
 \begin{equation}		
 \hat{g}_{Cl,ij}(t) = J_{EI}\sum_{i_0=-k_I}^{k_I}\sum_{j_0=-k_I}^{k_I} F_{i-i_0,j-j_0}K^{\mathrm{Loc,I}}_{i_0+k_I,j_0+k_I} 
+ J_{In}U 
\end{equation} 

 In this work, $J_{EI} = 250$, $J_{In} = 50$, which are determined by the condition  $(1-\gamma)/\gamma = 5$ when $\gamma = 1/6 $ based on phenomenological considerations\cite{liou2017multivariate}

The $k_E$ and $k_I$ are the radius of the nearest neighborhood in the two-dimensional neural network. Here, $k_E = 9$ and $k_I = 15$, indicating that recurrent inhibition operates over a broader spatial range than recurrent self-excitation. The $i_0$ and $j_0$, taken from either the excitatory or inhibitory subnetwork, denote the indices of neighboring neurons relative to position $(i,j)$. The scaling parameters $J_{EE}$, $J_{EI}$, and $J_{In}$ regulate the degree of self-excitation, as well as the localized and non-localized components of recurrent inhibition, respectively.   Thus, the synaptic conductances $g_{\mathrm{sE},ij}$ and $g_{\mathrm{Cl},ij}$ are updated according to first-order kinetics as follows:
 
{
	\small{
		\begin{subequations}
			\begin{align}  
				\frac{dg_{sE,ij}}{dt} &= -\frac{g_{sE,ij}}{\tau_E} + \overline{g}_{\mathrm{sE}}\cdot\hat{g}_{E,ij}(t)
				\label{gsyE}\\
				\frac{dg_{Cl,ij}}{dt} &= -\frac{g_{Cl,ij}}{\tau_I} +  \overline{g}_{\mathrm{Cl}}\cdot\hat{g}_{Cl,ij}(t) \label{gsyI} 
			\end{align}
			\label{EQ3}
\end{subequations}}}
 Here, $\tau_E$ and $\tau_I$ denote the decay time constants of excitatory and inhibitory synapses, respectively, with $\tau_E = \tau_I = 15$ ms. The parameters $\overline{g}_{\mathrm{sE}}$ and $\overline{g}_{\mathrm{Cl}}$ are normalized synaptic conductances, set to $\overline{g}_{\mathrm{sE}} = \overline{g}_{\mathrm{Cl}} = 1.0$ nS. 

\subsection{Statistical method for detecting focal seizures stages by the interspike interval (ISI) of firing rate}

\begin{figure}[htp]	
	\includegraphics[width=\linewidth]{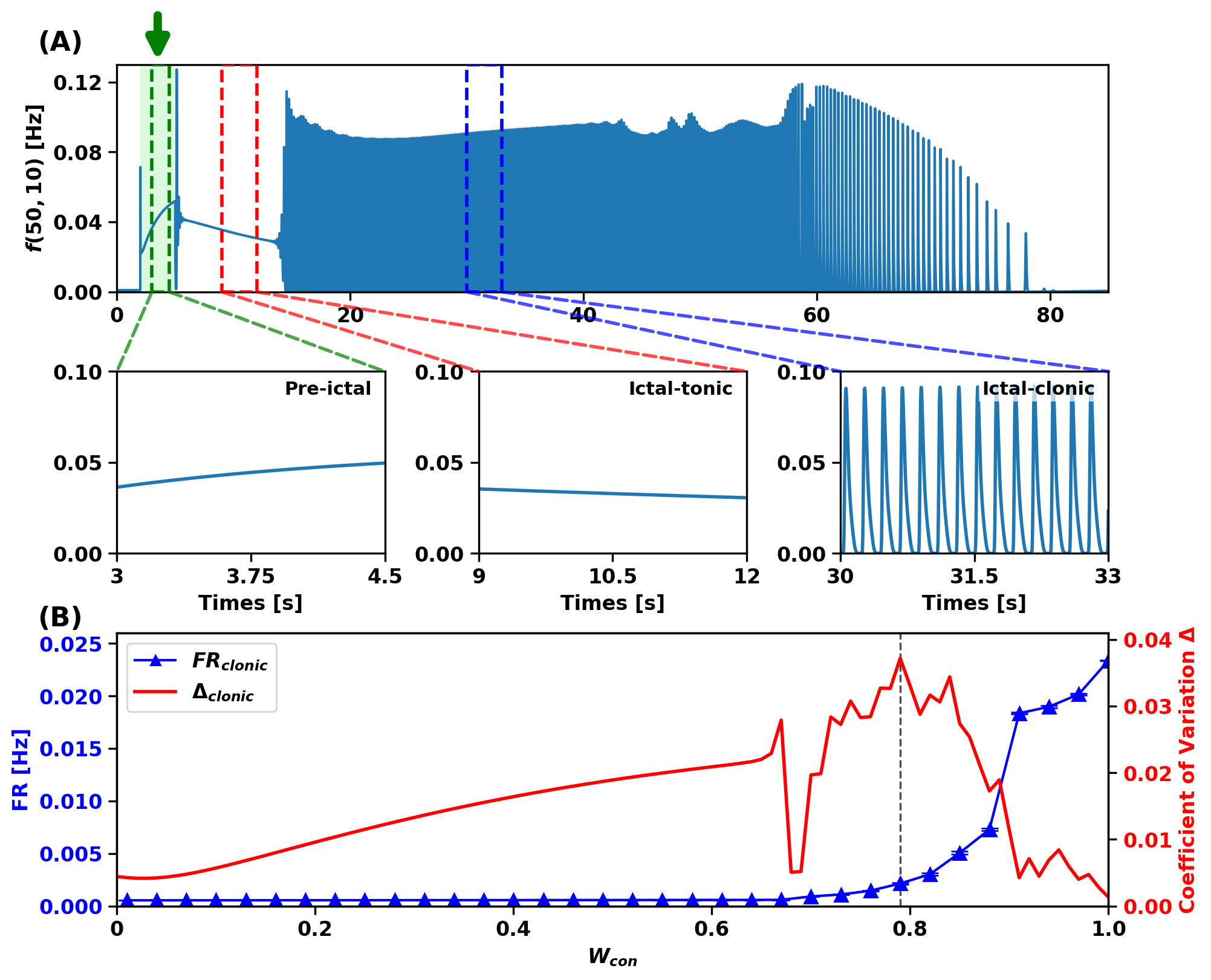}	  
	\caption{(A). Focal seizure dynamics across different stages; (B). Seizure stage transitions from failed seizure initiation to the clonic seizure characterized by statistical methods of mean firing rate ($\text{FR}$) and variability $\Delta$.}
	\label{fig:statis}
\end{figure}

Fig.\hyperref[fig:statis]{\ref{fig:statis}A} illustrates that the neural network exhibits distinct firing patterns across focal seizure stages as measured by the each neuronal firing rate $f_{i}(t)$ , defined in Eq.\hyperref[fvij]{\ref{fvij}}. Following a brief external excitation between 2000 and 5000 ms, the network enters the ictal-tonic stage at $T_{\mathrm{its}} = 5000$ ms, with the pre-ictal window defined as $T_{\mathrm{pre}} = 5000 - 2000$ ms. During the ictal-tonic stage, the interspike interval (ISI) of the firing rate exceeds 200 ms, whereas it remains close to 200 ms in the ictal-clonic stage. Accordingly, ISI is used to distinguish seizure stages, with the transition to the ictal-clonic stage ($T_{\mathrm{ics}}$) identified with $\text{ISI}= 2000$ ms (see Appendix B). The time windows for computing the mean firing rate during the ictal-tonic and ictal-clonic stages are defined as:

{\small{
\begin{equation}
	T_{tonic}=\left\{
	\begin{array}{rcl}
		\mathrm{T_{ics}}-\mathrm{T_{its}},  & &   \text{With ictal-tonic}\\
		0, & & \text{Without ictal-tonic}
	\end{array} \right.
	\label{Eq:TW2}
\end{equation}}}

{\small{
\begin{equation}
	T_{clonic}=\left\{
	\begin{array}{rcl}
		(\mathrm{T_{ics}}+30000)-\mathrm{T_{ics}},  & &   \text{With ictal-tonic}\\
		35000-5000, & & \text{Without ictal-tonic}
	\end{array} \right.
	\label{Eq:TW3}
\end{equation}}}
The presence or absence of the ictal-tonic stage is indicated by  \hyperref[Eq:TW2]{Eq.(\ref{Eq:TW2})} and \hyperref[Eq:TW3]{Eq.(\ref{Eq:TW3})}, respectively, which determine whether the focal seizure evolution includes or excludes the ictal-tonic stage. To capture the seizure dynamics across the entire excitatory population, the stages of global dynamical behavior can be measured using the well-known mean firing rate. The mean firing rate, $\mathrm{FR}$, is calculated following this procedure: First, for each time window $T_P$, the time-average of firing rate  $f_i(t)$  in trial $i$ for each seizure stage is defined as:
{\small{
\begin{equation}
	fr_i = \left<f_i(t)\right>_{T_P} = \frac{1}{T_{P}}\int_{0}^{T_{P}} f_i(t) \,dt  \quad P \in\text{(pre, tonic, clonic)}
\end{equation}}}

 We then further compute the trial-averaged  $fr_i$ across selected trials with seizure dynamics:

{\small{
\begin{equation}
	\mathrm{FR} = \left<fr_i\right> = \frac{\sum_{i=1}^{N_{tr}} \left<f_i(t)\right>_{T_P}}{N_{tr}} =  \frac{1}{N_{tr}}\sum_{i=1}^{N_{tr}}fr_i
\end{equation}}}

Where $N_{tr} = 25$ denotes the total number of trials.The rationale for selecting this specific local population is that these units are located in the core region of the seizure, allowing them to fully capture the entire process of seizure initiation, propagation, and termination. Motivated by the critical brain hypothesis, we also introduce the variability $\Delta$ to numerically determine the critical point of $W_{con}$, with $\Delta$ defined as\cite{liu2025recovery}:

{\small{
\begin{equation}
	\Delta = \dfrac{\sqrt{\left<fr_i^2\right>-\left<fr_i\right>^2}}{\left<fr_i\right>} = \dfrac{\sqrt{{\left<fr_i^2\right>}- \mathrm{FR}^2 }}{\mathrm{FR}}
	\label{Eq.varia}
\end{equation}}}

Here $\left< fr_i^2 \right> = \frac{1}{N_{tr}} \sum_{i=1}^{N_{tr}} fr_i^2$. Variability $\Delta$ represents the fluctuations, with the maximum value of $\Delta$ identifying the critical point between different states.  A larger $\Delta$ indicates greater variability in $fr_{i}$ across  selected neurons, whereas a smaller $\Delta$ indicates that  $fr_{i}$ values are tightly clustered around the mean firing rate FR. Therefore, variability $\Delta$ can be used to detect transitions between regions of seizure stages.

To interpret how seizure stages can be detected and quantified through these measurements, Fig.\hyperref[fig:statis]{\ref{fig:statis}B} illustrates a example of the emergence of ictal-clonic stages as the fraction parameter $W_{con}$ increases. The neural network fails to produce ictal-clonic stages when $FR = 0$ (blue line) and $\Delta >0$.  As $W_{con}$ increases, the neural network exhibits ictal-clonic firing activity, detectable through the mean firing rate when $\text{FR} > 0$ with the peak of $\Delta$ near 0.04 marking the transition from failed seizure initiation to the clonic stage (as detailed thereafter). Thus, these statistical measurements enable us to investigate the effects of chloride concentration on the emergence of seizure stages and its role in modulating seizure transitions.

\section{Results}

\subsection{Impact of intracellular chloride on seizure stage transitions in an excitatory–inhibitory network ($W_{con} = 1$)}

Neural networks composed of excitatory and inhibitory neurons connected via chemical synapses can reproduce distinct stages of focal seizure evolution, governed by intracellular chloride concentration and shifts in EI balance induced by local neuronal firing. Fig.\hyperref[fig:varWcon]{\ref{fig:varWcon}A} (also see Fig.\hyperref[fig:statis]{\ref{fig:statis}A}) illustrates focal seizure stages and their temporal transitions via the population's firing rate $f$. Specifically, following a brief excitatory stimulus during the pre-ictal stage (also see green shaded region in Fig.\hyperref[fig:statis]{\ref{fig:statis}A}), the network enters the ictal-tonic stage, characterized by a relatively stable firing rate (also see middle panel in Fig.\hyperref[fig:statis]{\ref{fig:statis}A}). A gradual decline in firing at the seizure then triggers repetitive bursting, marking the transition to the ictal-clonic stage (also right panel in Fig.\hyperref[fig:statis]{\ref{fig:statis}A}). As the ictal wavefront dissipates, the pre-termination stage emerges, with weaker and sparser bursts, before abrupt seizure termination occurs once the final burst propagates across the network. These transitions reflect dynamic shifts in excitation and inhibition, mediated by spatial summation and chloride-dependent modulation of the reversal potential at different excitation levels. These mechanisms were thoroughly examined in Ref.\cite{liou2020model} for the case $W_{con} = 1$.

The results in Fig.\hyperref[fig:statis]{\ref{fig:statis}A} and Fig.\hyperref[fig:varWcon]{\ref{fig:varWcon}A} indicate that EI balance is a key factor in regulating excitatory activity through spatial summation and chloride dynamics. Intracellular chloride concentration is governed by two opposing processes: influx and clearance via first-order kinetics [shown in Eq.\hyperref[chlorideflow]{(\ref{chlorideflow})}]\cite{kaila2014cation,doyon2016chloride}. These observations highlight a central question in focal seizure dynamics: how can the progression through different seizure stages be controlled? To explore this, we first assess the role of the parameter $W_{con}$ [Eq.\hyperref[chlorideflow]{(\ref{chlorideflow})}], which modulates chloride influx, thereby changing intracellular chloride levels. Given that synaptic weights also shape seizure evolution, we further investigate the combined effects of chloride influx and synaptic connectivity on seizure dynamics across stages.

\subsection{ Focal seizure dynamics and their transitions are controlled byintracellular chloride with variation of $W_{con}$}

\begin{figure}[htp]	
	\includegraphics[width=\linewidth]{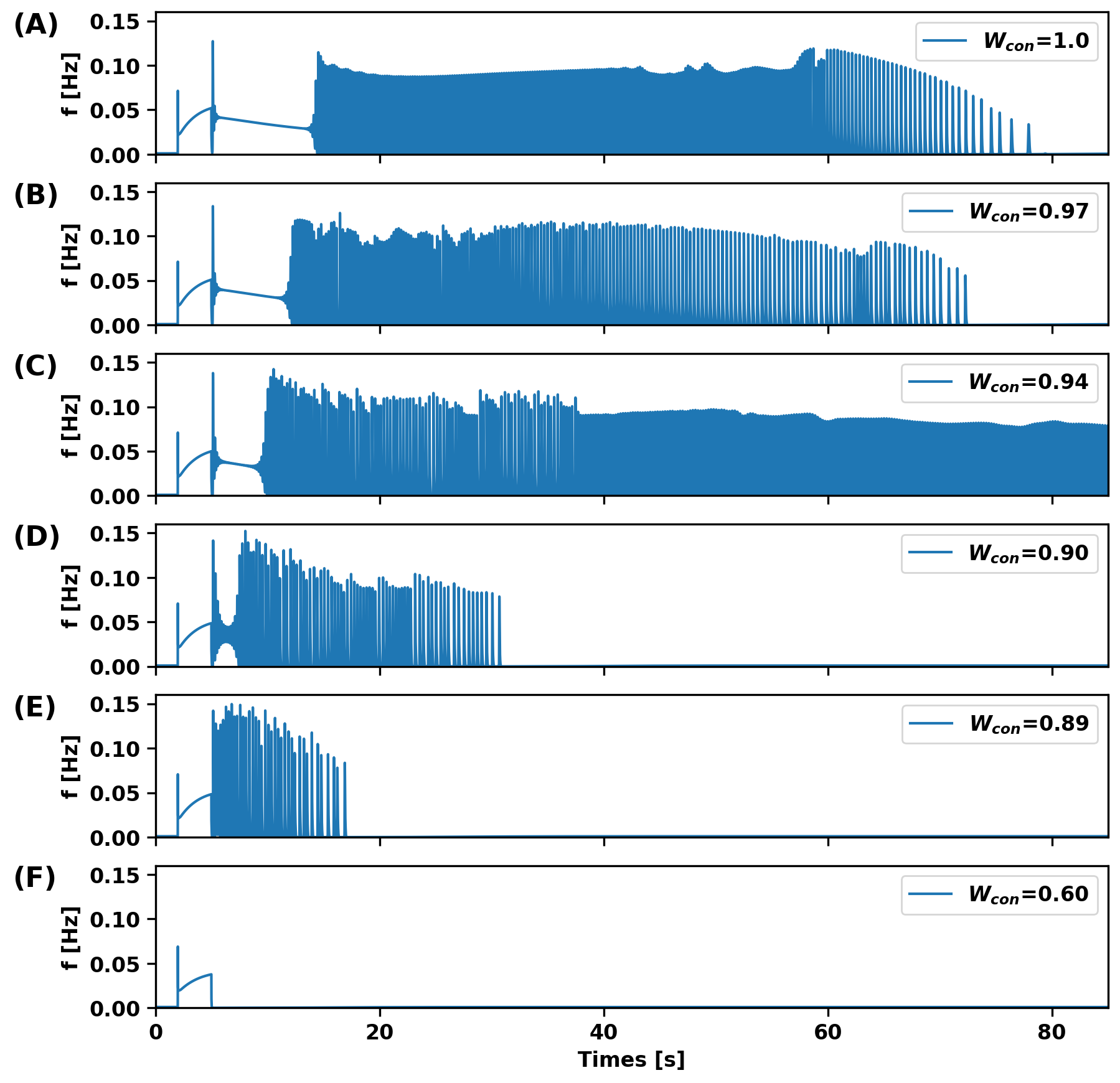}	  
	\caption{Stages of focal seizure dynamics across different fractions of $W_{con}$ at location (50,10):(A and B) Pre-ictal, ictal-tonic, and ictal-clonic.(C) Pre-ictal, ictal-tonic, and ictal-clonic in spiral waves.(D) Pre-ictal, ictal-tonic, and ictal-clonic (short time range).(E) Pre-ictal and ictal-clonic stages.(F) Failed seizure initiation.}
	\label{fig:varWcon}
\end{figure}

Fig.\hyperref[fig:varWcon]{\ref{fig:varWcon}} shows the spatiotemporal evolution of focal seizure stages at different values of $W_{con}$. A slight reduction to 0.97 (panel B) relative to $W_{con} = 1.0$ (panel A) narrows the ictal-tonic stage and leads to earlier termination during the ictal-clonic stage, though all characteristic stages remain present. At $W_{con} = 0.94$ (panel C), seizures persist in a prolonged ictal-clonic state as the network enters a spiral-wave regime. Further reduction to 0.90 (panel D)  shortens both ictal-tonic and ictal-clonic stages of the tonic–clonic seizure. These findings suggest that $W_{con}$ may play a key role in controlling focal seizure dynamics. Supporting this, at $W_{con} = 0.89$ (panel E), the network bypasses the ictal-tonic stage and transitions directly from pre-ictal to ictal-clonic activity, corresponding to the emergence of clonic seizures. At $W_{con} = 0.60$ (panel F), external excitation fails to trigger seizures. These results demonstrate that chloride influx, regulated by $W_{con}$, critically shapes seizure dynamics. However, the mechanistic link between $W_{con}$, intracellular chloride concentration, and seizure control remains to be fully resolved.

\begin{figure}[htp]	
	\includegraphics[width=\linewidth]{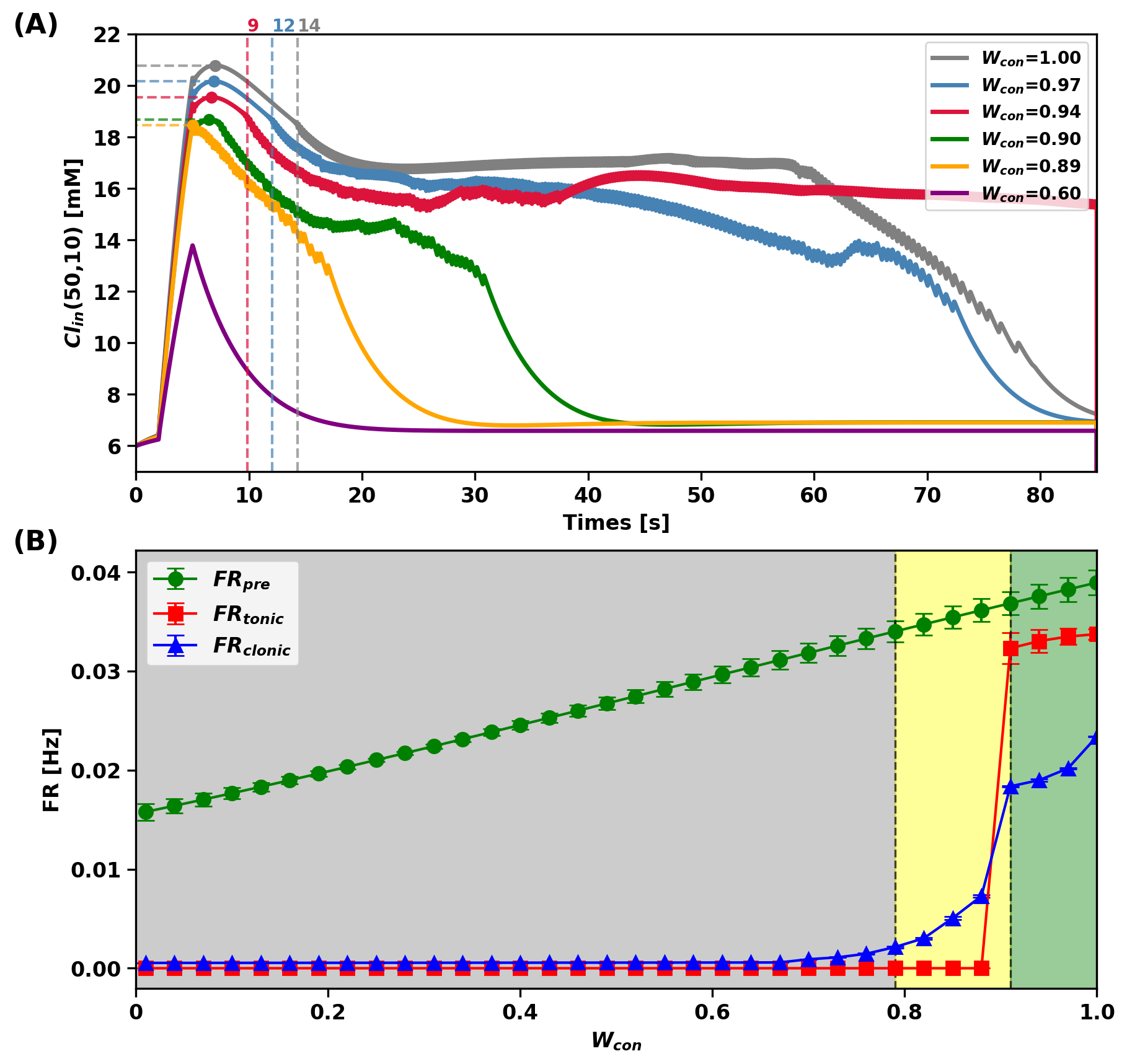}	  
	\caption{Effect of chloride influx on seizure dynamics with varying $W_{con}$. (A) Intracellular chloride concentration evolution. (B)Seizure stages and transitions across different focal seizure types. Gray denotes seizure offset, yellow indicates clonic seizures, and green represents tonic–clonic seizures, consistent with the color scheme used in subsequent figures.}
	\label{fig:chloride}
\end{figure}

We now examine how $W_{con}$ regulates chloride influx and thereby controls focal seizures.Fig.\hyperref[fig:chloride]{\ref{fig:chloride}A} shows intracellular chloride dynamics at location $(50,10)$ for the $W_{con}$ values used in Fig.\hyperref[fig:varWcon]{\ref{fig:varWcon}}. For $W_{con} = 1$ and $0.97$, external excitation produces a sharp rise in chloride ($\text{Cl}_{in} \geq 20$), followed by decay to moderately elevated levels with time constants of $\sim14$ s and $\sim12$ s, respectively. At $W_{con} = 1$, chloride remains elevated until $\sim60$ s, whereas at $0.97$ it declines more slowly, sustaining $\text{Cl}_{in} \geq 15$ until seizure termination. These dynamics correspond to sequential ictal-clonic, late ictal-clonic, pre-termination, and termination stages (Figs.\hyperref[fig:varWcon]{\ref{fig:varWcon}}A,B). At $W_{con} = 0.94$, chloride stabilizes at moderately elevated levels, supporting a spiral-wave state. At $W_{con} = 0.90$, it peaks slightly above 18 and returns to baseline by 40 s, while at $0.89$ it decays to baseline within 30 s, leading to early termination . For $W_{con} = 0.60$, chloride peaks below 14 and rapidly returns to baseline within 20 s .

Fig.\hyperref[fig:chloride]{\ref{fig:chloride}B} shows how the mean firing rate varies with $W_{con}$, with seizure type transitions identified by variability $\Delta$ (Eq.\hyperref[Eq.varia]{\ref{Eq.varia}}). For $W_{con} \leq 0.79$ (gray region), only the pre-ictal stage persists ($\text{FR}_{pre} > 0.01$; $\text{FR}_{tonic} = \text{FR}_{clonic} = 0$), indicating seizure termination (see Fig.\hyperref[fig:varWcon]{\ref{fig:varWcon}}F). In the range $(0.79,0.91]$ (yellow region), the ictal-clonic mean firing rate increases while the ictal-tonic rate remains zero, producing sequential pre-ictal and ictal-clonic stages (see Fig.\hyperref[fig:varWcon]{\ref{fig:varWcon}}E). For $W_{con} > 0.91$ (green region), all three stages -- pre-ictal, ictal-tonic, and ictal-clonic -- emerge, consistent with Figs.\hyperref[fig:varWcon]{\ref{fig:varWcon}A-C}. Together, Figs.\hyperref[fig:varWcon]{\ref{fig:varWcon}} and \hyperref[fig:chloride]{\ref{fig:chloride}} illustrate seizure dynamics and chloride influx of the excitatory population across $W_{con}$. The underlying mechanism responsible for these observations is the depolarizing shift in the chloride reversal potential, which arises from chloride homeostasis regulated by variations in $W_{con}$ (see Appendix C). 

\begin{figure}[htp]	
	\includegraphics[width=\linewidth]{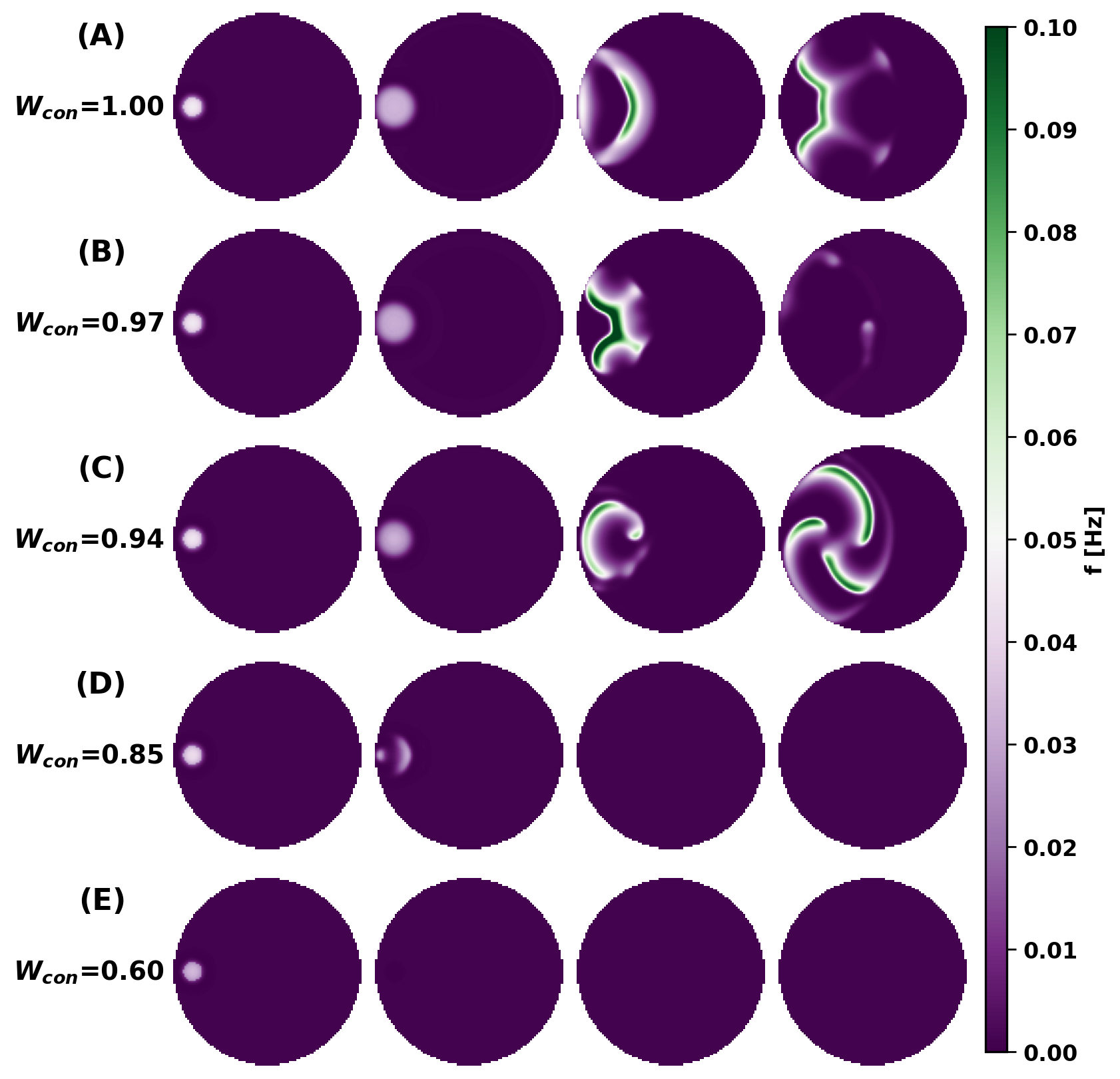}	  
	\caption{Spatiotemporal dynamics of focal seizures in the entire excitatory population as the $W_{con}$ varies.}
	\label{fig:spatioPatt}
\end{figure}

 To better illustrate the emergence of collective dynamics across the neural network, Fig.\hyperref[fig:spatioPatt]{\ref{fig:spatioPatt}} presents spatiotemporal patterns of status epilepticus, spiral waves, and seizure termination for different $W_{con}$ values. Specifically, the neural network exhibits sequential ictal-clonic, late ictal-clonic, pre-termination, and termination firing patterns when $W_{con} = 1.0$ and $W_{con} = 0.97$, consistent with Figs.\hyperref[fig:varWcon]{\ref{fig:varWcon}A,B}. At $W_{con} = 0.94$, the emergence of spiral waves, shown in the last panel of Fig.\hyperref[fig:spatioPatt]{\ref{fig:spatioPatt}C}, gives rise to prolonged ictal–clonic activity without seizure termination. These spiral waves progressively dominate the spatial domain, become self-sustaining, and persist indefinitely,as quantified by the spiral-wave centers (see Appendix C), resulting in a model representation of status epilepticus. Spiral-wave dynamics have been observed in both animal experiments\cite{huang2010spiral} and humans, suggesting that spiral-wave dynamics may serve as a marker of increased risk for status epilepticus\cite{liou2020model}. As $W_{con}$ decreases further, focal seizures are successfully controlled, as observed at $W_{con} = 0.85$ and $W_{con} = 0.60$. These findings from the entire neural network reinforce that focal seizures can be effectively regulated by reducing $W_{con}$, which affects chloride influx.

\subsection{Focal seizure dynamics and their transitions are regulated by excitatory and inhibitory balance}

\begin{figure}[htp]	
	\includegraphics[width=\linewidth]{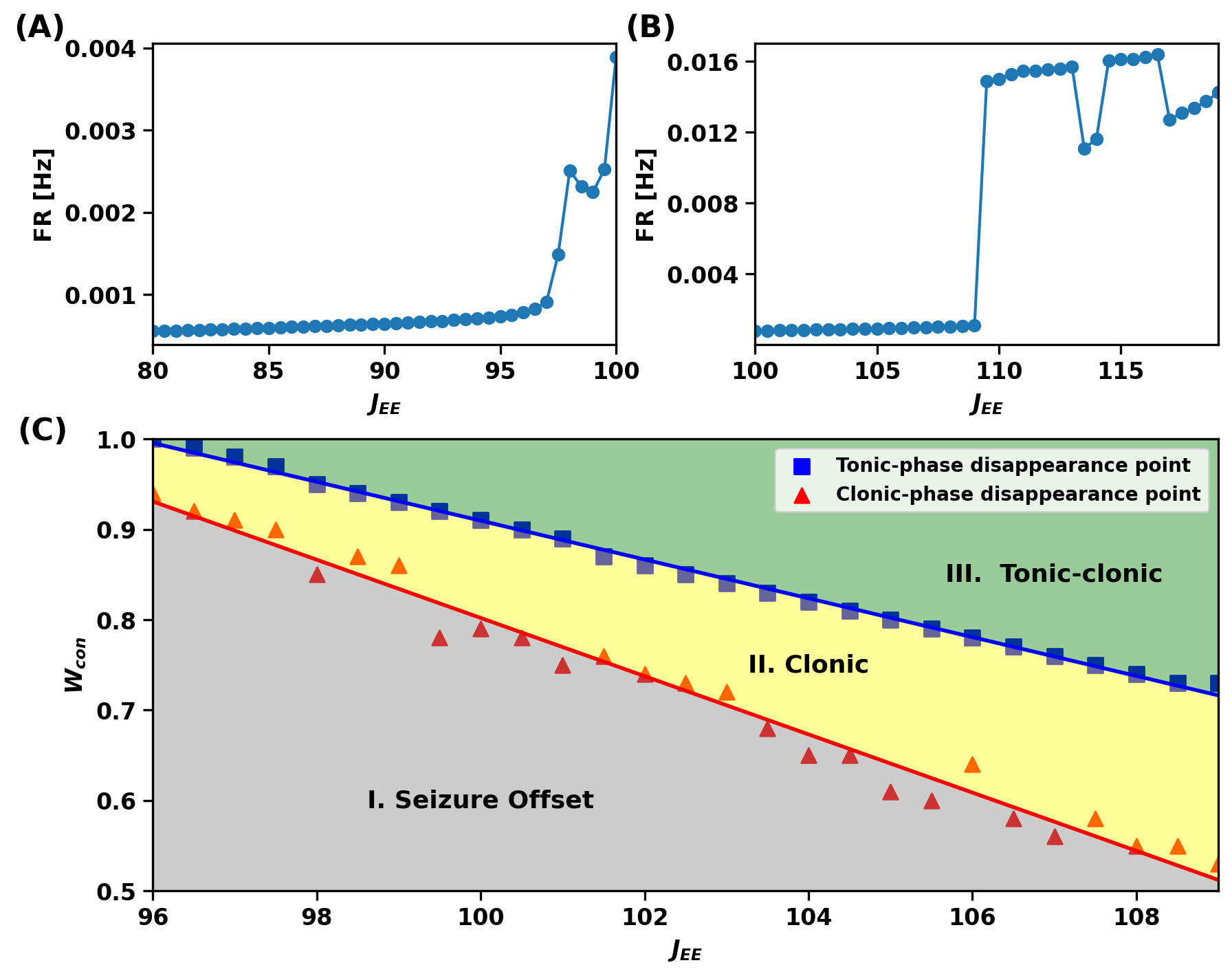}	  
	\caption{Effect of recurrent excitation and intracellular chloride on seizure dynamics. (A,B) The  recurrent excitation  $J_{EE}$ regulates seizure onset with $J_{EI}=250$, $W_{con}=1.0$. (C) Seizure types in the $J_{EE}/W_{con}$ space. }
	\label{fig:JEEvary}
\end{figure}

The results from Fig.\hyperref[fig:statis]{\ref{fig:statis}} to Fig.\hyperref[fig:spatioPatt]{\ref{fig:spatioPatt}} characterize focal seizure dynamics under fixed excitation ($J_{EE}$) and inhibition ($J_{EI}$). We next investigate how these dynamics and their transitions are regulated by the excitatory–inhibitory balance across $W_{con}$, which modulates intracellular chloride influx. Fig.\hyperref[fig:JEEvary]{\ref{fig:JEEvary}} illustrates the mean firing rate in the $J_{EE}/W_{con}$ parameter space. When excitatory coupling is weak ($J_{EE} \le 96$, Fig.\hyperref[fig:JEEvary]{\ref{fig:JEEvary}A}), external excitatory input fails to evoke focal seizure activity, resulting in $\text{FR} = 0$ Hz. Conversely, when excitatory coupling is strong ($J_{EE} \ge 109$, Fig.\hyperref[fig:JEEvary]{\ref{fig:JEEvary}B}), the network exhibits spontaneous high-frequency activity even in the absence of external input, with $\text{FR} \ge 0.012$ Hz. These findings indicate that the excitation–inhibition balance critically governs seizure onset, consistent with experimental evidence that cortical hyperexcitability destabilizes network dynamics.

To further examine this mechanism, Fig.\hyperref[fig:JEEvary]{\ref{fig:JEEvary}C} maps focal seizure dynamics and transitions across the $J_{EE}/W_{con}$ parameter space, with $J_{EE}$ ranging from 96 to 109. When both $J_{EE}$ and $W_{con}$ lie within the gray region, the network remains in failed seizure initiation and fails to propagate seizures, reflecting a stable but subthreshold regime reminiscent of clinically observed pre-seizure activity. In the light yellow region, seizures transition directly from the pre-ictal to the ictal–clonic stage, bypassing tonic activity analogous to sudden-onset seizures without prodromal symptoms\cite{noachtar2009semiology,thijs2021autonomic}. The green region corresponds to the full progression through pre-ictal, ictal–tonic, and ictal–clonic stages, closely matching the temporal evolution of clinical tonic–clonic seizures\cite{thijs2021autonomic}.

\begin{figure}[htp]	
	\includegraphics[width=\linewidth]{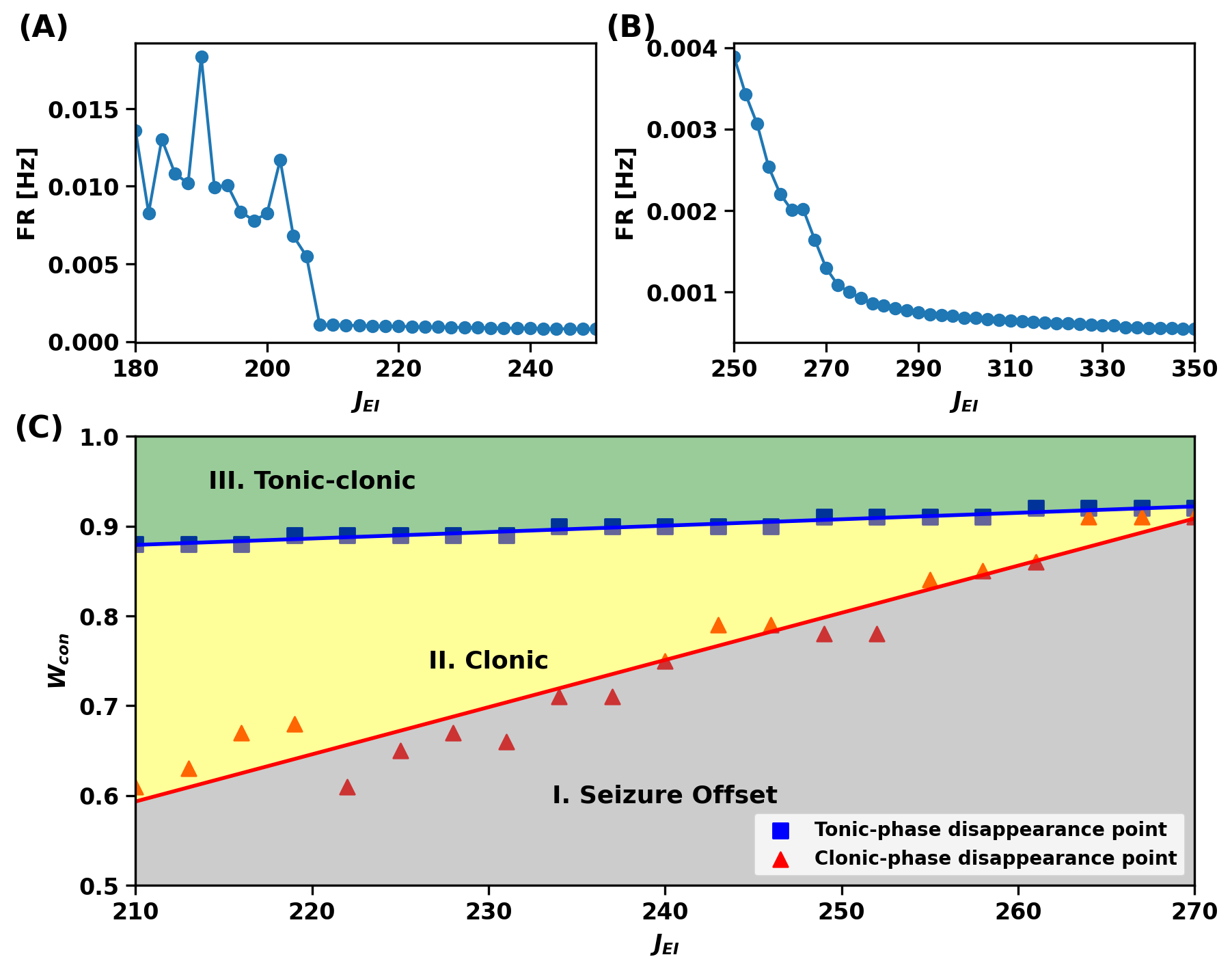}	  
	\caption{Effect of recurrent inhibition  and intracellular chloride  on seizure dynamics. (A,B) The recurrent inhibition  $J_{EI}$ regulates seizure onset with $J_{EE}=100$, $W_{con}=1.0$. (C) Seizure types in the $J_{EI}/W_{con}$ space. }
	\label{fig:JEIvary}
\end{figure}

When $W_{con}$ is reduced (e.g., $W_{con}=0.5$ in Fig.\hyperref[fig:JEEvary]{\ref{fig:JEEvary}C}), seizure activity is more readily suppressed, as indicated by the expanded gray region. This suggests that weaker chloride conductance may mitigate hyperexcitability by preventing inhibitory collapse. As $W_{con}$ increases, however, the seizure offset regime narrows, and the network more easily transitions into clonic seizure or tonic-clonic seizure, reflecting the destabilizing influence of impaired chloride homeostasis. Likewise, increasing $J_{EE}$ enlarges the green region of tonic–clonic seizures, indicating that excessive excitation broadens the parameter range supporting seizure-like dynamics.  A decrease in  $W_{con}$ enhances inhibition, whereas an increase in $J_{EE}$ strengthens excitation, together maintaining the excitation–inhibition balance in the seizure-offset region and thereby suppressing the onset of focal seizures. 

For comparison with Fig.\hyperref[fig:JEEvary]{\ref{fig:JEEvary}}, Fig.\hyperref[fig:JEIvary]{\ref{fig:JEIvary}} illustrates focal seizure dynamics as a function of the excitatory–inhibitory balance by systematically varying the inhibitory coupling strength ($J_{EI}$). Under weak inhibition ($J_{EI} \le 210$, Fig.\hyperref[fig:JEIvary]{\ref{fig:JEIvary}A}), the network exhibits spontaneous high-frequency activity even without external excitatory input, with $\text{FR} \ge 0.005$ Hz. In contrast, strong inhibition ($J_{EI} \ge 270$, Fig.\hyperref[fig:JEIvary]{\ref{fig:JEIvary}B}) suppresses seizure generation, with $\text{FR} = 0$ Hz. These findings also underscore the pivotal role of excitation–inhibition balance in governing seizure onset, consistent with experimental evidence that cortical hyperexcitability destabilizes network dynamics.

Fig.\hyperref[fig:JEIvary]{\ref{fig:JEIvary}C} maps seizure types across the $J_{EI}/W_{con}$ phase space ($J_{EI} \in [210, 270]$). When $W_{con}$ lies below the blue transition line separating the light yellow and green regions, stronger inhibition expands the seizure offset region (light gray) while reducing the clonic region (light yellow), where seizures transition directly from pre-ictal to clonic stages. This observation aligns with experimental evidence that enhanced inhibition stabilizes network activity and delays seizure onset. At higher $W_{con}$ values (light green region), however, seizure dynamics become largely insensitive to inhibitory strength, consistently exhibiting tonic–clonic patterns. This insensitivity reflects the destabilizing effect of impaired chloride homeostasis: elevated $W_{con}$ diminishes GABAergic efficacy and promotes hyperexcitability\cite{cherubini2022dysregulation,tang2021role}.

\begin{figure}[htp]	
	\includegraphics[width=\linewidth]{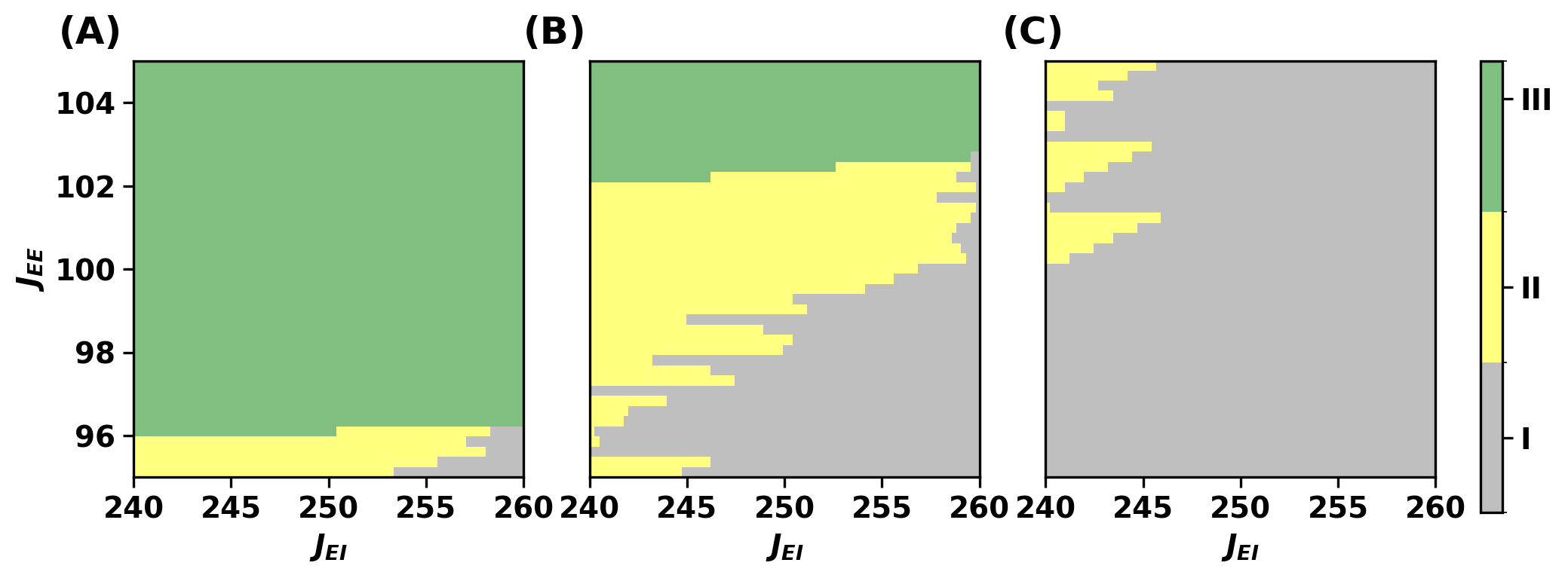}	  
	\caption{Effect of recurrent inhibition ($J_{EI}$) and excitation ($J_{EE}$) on seizure dynamics across different values of $W_{con}$, where (A) $W_{con}=1.0$, (B) $W_{con}=0.85$ and (C) $W_{con}=0.6$.}
	\label{fig:JEIJEE}
\end{figure}

We provide a broader perspective on focal seizure dynamics in the $J_{EI}/J_{EE}$ parameter space across different values of $W_{con}$. When $W_{con} = 1.0$, Fig.\hyperref[fig:JEIJEE]{\ref{fig:JEIJEE}A} shows that the network generates sequential pre-ictal, ictal-tonic, and ictal-clonic stages under strong excitation (green region), suggesting that cortical hyperexcitability destabilizes network dynamics, leading to focal seizures. In contrast, under weaker excitation, the network exhibits only pre-ictal and ictal-clonic stages, with no distinct ictal-tonic stage (see Figs.\hyperref[fig:varWcon]{\ref{fig:varWcon}}–\hyperref[fig:spatioPatt]{\ref{fig:spatioPatt}}) as $J_{EI}$ increases but remains moderate (yellow region). When inhibition is sufficiently strong, focal seizures are effectively suppressed and terminate rapidly following external stimulation (gray region). Reducing $W_{con}$ to $0.85$ and $0.6$ (compared with Fig.\hyperref[fig:JEIJEE]{\ref{fig:JEIJEE}A}) enlarges the stable regions (deep gray region in Figs.\hyperref[fig:JEIJEE]{\ref{fig:JEIJEE}B,C}), indicating that seizures are more easily terminated at lower $W_{con}$ values. Overall, these results demonstrate that while strengthening inhibition can suppress seizure propagation under normal chloride regulation, chloride dysregulation can override inhibitory control, driving robust tonic-clonic seizures, consistent with clinical and experimental observations\cite{glykys2017chloride}.

\subsection{The underlying mechanisms of focal seizure types and transitions driven by EI-balance shifts}\label{mechan}

In Eq.\hyperref[Menpo]{(\ref{Menpo})},  $I_{sE,ij}$ and $I_{Cl,ij}$ are referred to as the excitatory and inhibitory post-synaptic currents (EPSCs and IPSCs), respectively. EPSCs drive neuronal depolarization, increasing the likelihood of action potential firing, whereas IPSCs induce neuronal hyperpolarization, reducing the likelihood of firing\cite{hille2001ion}. Three  primarily factors influence post-synaptic currents: synaptic conductance ($W_{con}$, the fraction of $\text{Cl}^-$ and $\mathrm{HCO_3^-}$ conductance), the reversal potential (regulated by $\text{Cl}_{in}$), and the synaptic weights between different neurons. Specifically, $W_{con}$  controls the fraction of $\mathrm{Cl}^-$  and  $\mathrm{HCO_3^-}$ conductances, thereby regulating intracellular chloride levels and modulating the chloride reversal potential $\text{Cl}_{in}$.

\textbf{Progressive depolarizing (excitatory) shifts of chloride reversal potential  $\text{Cl}_{in}$ reduces inhibition and promotes excitation.} A characteristic feature of inhibitory synapses is that the synaptic reversal potential $E_{syn}$ typically lies in the range of $-70$ to $-75$ mV\cite{gerstner2014neuronal}. However, the chloride reversal potential $E_{Cl}$, defined in Eq.\hyperref[ECL]{(\ref{ECL})}, which governs GABAergic synaptic inhibition, is primarily determined by the transmembrane chloride ion concentration gradient\cite{raimondo2017neuronal,currin2022computational,kaila2014cation}.

\begin{figure}[htp]	
	\includegraphics[width=\linewidth]{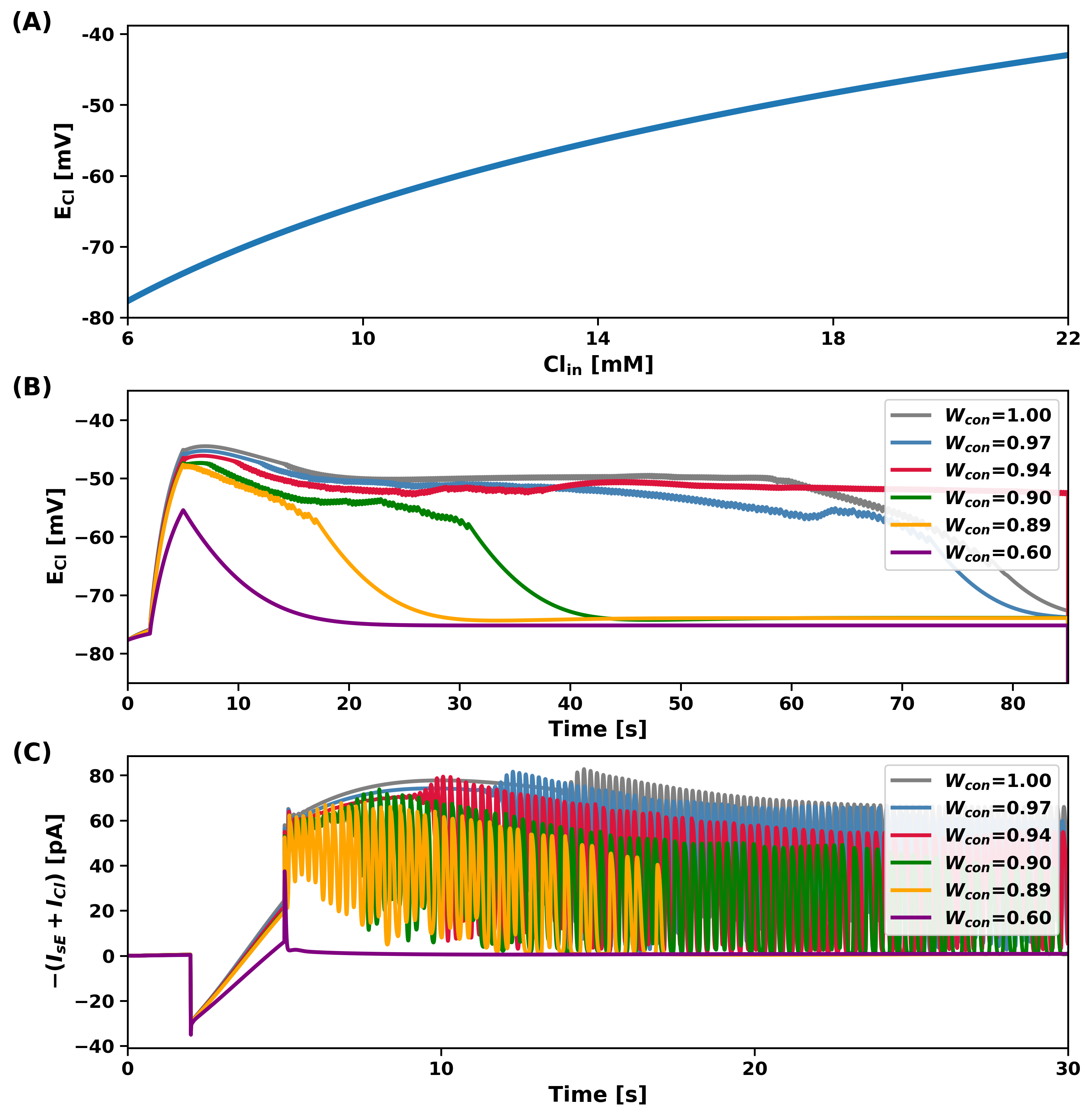}	  
	\caption{The progressive depolarizing (excitatory) shifts in the GABAergic reversal potential resulting from altered chloride homeostasis. (A) Reversal potential as a function of chloride homeostasis. (B) Evolution of the reversal potential and synaptic currents  (C) over time at different values of$W_{con}$. }
	\label{fig:mech}
\end{figure}

To better illustrate how the reversal potential $\text{Cl}_{in}$ depends on intracellular chloride concentration, Fig.\hyperref[fig:mech]{\ref{fig:mech}} shows progressive depolarizing shifts of the GABAergic reversal potential resulting from altered chloride homeostasis. Specifically, Fig.\hyperref[fig:mech]{\ref{fig:mech}A} shows that the reversal potential $E_{Cl}$  increases gradually from -78 mV (calculated using Eq.\hyperref[ECL]{(\ref{ECL})} for $\text{Cl}_{in} = 6$) to -41 mV (for $\text{Cl}_{in} = 22$) as $\text{Cl}{in}$ rises. Here, $\text{Cl}_{in} = 22$ corresponds to the peak intracellular chloride concentration observed at $W_{con} = 1.0$, as shown in Fig.\hyperref[fig:chloride]{\ref{fig:chloride}A}. These results suggest that an increase in the reversal potential reflects a gradual loss of inhibitory efficacy, as neurons become less hyperpolarized (inhibition), a phenomenon commonly observed in epilepsy or following neuronal injury and associated with increased network excitability and seizure susceptibility. Fig.\hyperref[fig:mech]{\ref{fig:mech}B} explores the evolution of the reversal potential, calculated from Eq.\hyperref[ECL]{(\ref{ECL})},suggesting that peak values of the reversal potential $E_{Cl}$ are decreased by stronger inhibition, achieved by reducing $W_{con}$. Fig.\hyperref[fig:mech]{\ref{fig:mech}C}further demonstrates that decreasing $W_{con}$ reduces synaptic currents, thereby enhancing synaptic inhibition. In other words, stronger synaptic inhibition (e.g., $W_{con} = 0.6$) suppresses the initiation of focal seizures, as observed in Figs.\hyperref[fig:varWcon]{\ref{fig:varWcon}}--\hyperref[fig:spatioPatt]{\ref{fig:spatioPatt}}. This constitutes a primary mechanism for controlling seizure types and their transitions, whereby progressive depolarizing shifts driven by increases in $\text{Cl}_{in}$ weaken synaptic inhibition, contributing to the onset of focal seizures.

\begin{figure*}[htp]	
	\includegraphics[width=\linewidth]{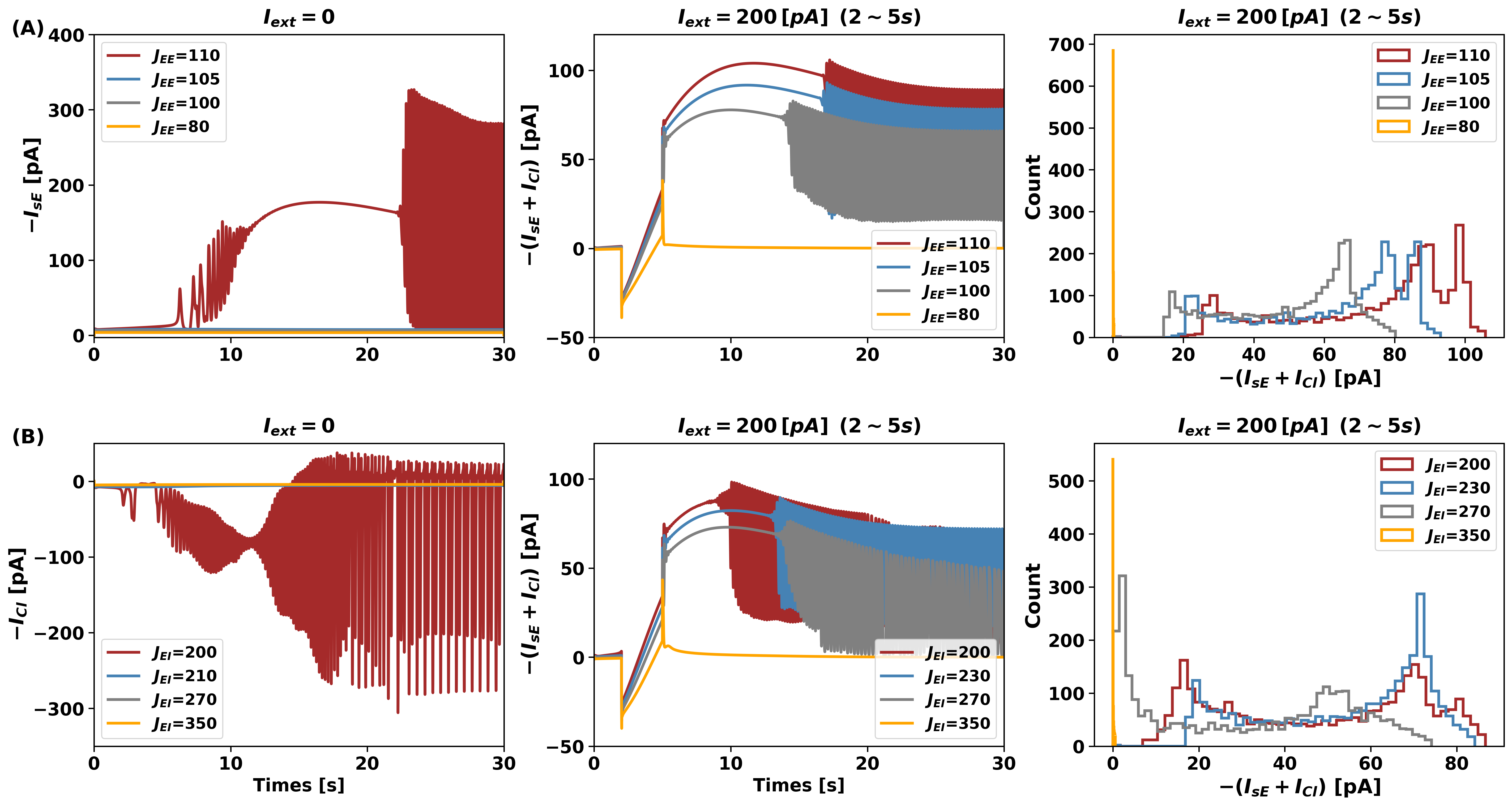}	   
	\caption{Evolution of post-synaptic currents at varying levels of excitation (A, scaled by $J_{EE}$) and inhibition  (B, scaled by $J_{EI}$). }
	\label{fig:jeevsjei}
\end{figure*}

\textbf{Balanced EPSCs and IPSCs are determined by synaptic weights.} The synaptic weights between neurons serve as the second mechanism that shifts the EI balance towards excitation, triggering the onset of focal seizures. To better explain the role of synaptic weights in the emergence of focal seizures, Fig.\hyperref[fig:jeevsjei]{\ref{fig:jeevsjei}} shows the postsynaptic currents at varying levels of excitation and inhibition, modulated by the parameters $J_{EE}$ and $J_{EI}$, respectively. As shown in Fig.\hyperref[fig:jeevsjei]{\ref{fig:jeevsjei}A}, both EPSCs (left panel, $I_{ext} = 0$) and the sum of EPSCs and IPSCs (middle panel, $I_{ext} = 200$ between 2 to 5 seconds) increase as $J_{EE}$ rises, shifting the balance towards excitation and promoting focal seizure onset. In contrast, under the same $I_{ext}$ setting for each panel, Fig.\hyperref[fig:jeevsjei]{\ref{fig:jeevsjei}B} demonstrates that both IPSCs (left panel) and the total of EPSCs and IPSCs (middle panel) decrease with higher $J_{EI}$, shifting the balance towards inhibition and preventing the onset of focal seizure. Histograms of postsynaptic currents (right panels in Fig.\hyperref[fig:jeevsjei]{\ref{fig:jeevsjei}}) further demonstrate that increasing $J_{EE}$ or decreasing $J_{EI}$ shifts the distributions toward higher current values, consistent with enhanced excitation.

In a nutshell, focal seizure dynamics and the transitions between seizure stages are regulated by the excitation–inhibition balance through two primary mechanisms: (i) depolarizing shifts in the chloride reversal potential arising from chloride homeostasis by variations in $W_{con}$, and (ii)modulation of synaptic weights via the excitatory and inhibitory coupling parameters $J_{EE}$ and $J_{EI}$.

\section{discussion}

Focal seizures recorded with intracranial electrodes in patients undergoing presurgical evaluation typically progress through distinct stages, most notably the ictal-tonic and ictal-clonic phases \cite{gentiletti2022focal}. In temporal lobe epilepsy, seizure onset is often marked by low-voltage fast activity in the gamma range (30–80 Hz) \cite{capitano2024preictal}, which evolves into irregular spiking patterns and periodic bursts. Over time, these bursts decelerate in frequency, characteristic of the ictal-clonic stage (Fig.\hyperref[fig:statis]{\ref{fig:statis}A}), before abruptly terminating \cite{liou2020model}. To explore the mechanisms underlying stage transitions during focal seizures, we developed an excitatory–inhibitory network model, reflecting the critical interplay between excitatory pyramidal neurons and inhibitory interneurons in shaping cortical dynamics.

Computational modeling provides an essential framework for integrating diverse data sources and elucidating the physiological mechanisms of epilepsy \cite{li2024dynamical}. Connectivity analyses using EEG and MEG reveal that functional network organization, including node degree and small-world topology, influences seizure onset, propagation, and frequency \cite{ guye2010graph}. Seizure dynamics have been modeled using both deterministic and stochastic approaches \cite{lytton2008computer}. Deterministic models span multiple scales, ranging from ion channel kinetics in single neurons \cite{hodgkin1952quantitative} to large-scale population dynamics \cite{deco2008dynamic}. In contrast, stochastic models account for variability in brain activity arising from noise at molecular, synaptic, and network levels, and explore how randomness shapes epileptic dynamics \cite{brari2022new, wang2023noise, feldman2024neurobiology}. Collectively, these observations highlight a critical link between coupling strength, connectivity, and node susceptibility in mediating the relationship between chloride homeostasis and neuronal hyperexcitability. However, the role of disrupted chloride extrusion in regulating EI balance, seizure activity, and stage transitions from fundamental biophysical mechanisms remains poorly understood. 

The classical view that seizures arise from an imbalance between synaptic excitation and inhibition provides only a partial explanation for the stages observed during focal seizures \cite{milligan2021epilepsy, untiet2024astrocyte}. While changes in excitatory and inhibitory drive are pivotal in shaping network excitability\cite{zhou2022short,liu2025formation,shan2025short}, this framework alone cannot fully explain the complex electrographic dynamics or the structured progression of temporal lobe epilepsy across ictal phases. Growing evidence further challenges the idea that synaptic mechanisms are the sole drivers of ictogenesis \cite{feng2003low, gentiletti2022focal}. For instance, seizure-like events persist even under suppressed synaptic transmission and can propagate across lesioned hippocampal regions, suggesting that classical synaptic signaling is not essential for seizure generation or spread \cite{lian2001propagation}. Moreover, light-induced neocortical seizures were associated with severe extracellular $\rm{Ca^{2+}}$ depletion, incompatible with conventional synaptic transmission, further supporting the notion that seizures can emerge independently of classical synaptic mechanisms \cite{pumain1985chemical, huang2021tresk}.

In this study, we assume inhibitory interneurons respond instantaneously and project to corresponding excitatory neurons at the same location for two reasons. First, the proposed neuronal network is modeled based on a rate-based framework, where the firing rate $f$ of each population in the neural field effectively replaces the activity of individual interneurons, whose isolated dynamics play a relatively limited role in regulating focal seizures. Second, we argue that explicitly modeling complex inhibitory interneuron dynamics is not required to replicate key seizure features addressed in this work. For example, propagation delays between excitatory and inhibitory populations do not alter focal dynamics in our model (Appendix D). Although this simplification does not diminish the well-established importance of interneurons in seizure evolution (see review \cite{Pelkey}), inhibitory interneurons remain crucial for seizure initiation and propagation \cite{gentiletti2022focal,Pelkey,Gnatkovsky2008,Tahra2017}. Various experimental and theoretical studies have shown that both elevated extracellular potassium and increased intracellular chloride contribute to depolarizing GABAergic responses\cite{gentiletti2022focal,Pelkey,Gnatkovsky2008,Tahra2017,gentiletti2022focal,Gnatkovsky2008}. In addition, inhibitory interneurons influence seizure initiation through extracellular potassium accumulation resulting from changes in spiking activity \cite{gentiletti2022focal,Gnatkovsky2008}.

These findings imply that modulation of synaptic gain alone cannot account for seizure initiation and progression. Instead, focal seizures likely engage additional cellular and network mechanisms, such as disrupted ion homeostasis, ephaptic interactions, gap junction-mediated coupling, and activity-dependent changes in intrinsic excitability \cite{de2018potassium}. Recognizing these non-synaptic factors broadens the framework for seizure generation and highlights the limitations of the classical excitation–inhibition imbalance model in explaining the electrographic and cellular features of epilepsy.

While seizures often progress through distinct stages, the mechanisms governing these transitions remain poorly understood. In this study, we introduce a control parameter for intracellular chloride influx ($W_{con}$) using a computational excitatory–inhibitory network model to investigate how chloride regulation and excitation–inhibition balance shape seizure activity. Our results show that chloride homeostasis is critical for determining whether seizures initiate, how they evolve, and which patterns predominate. Disruption of $W_{con}$ alters seizure progression, leading to shortened clonic activity, direct transitions into clonic seizures, or robust tonic–clonic dynamics that become insensitive to inhibitory strength. These findings offer mechanistic insights into how impaired chloride regulation contributes to cortical hyperexcitability and network instability, advancing our understanding of seizure generation and progression.

\begin{acknowledgements}
The authors thank all members of the HQU–BUAS–JSU–NBU–BNU–JXU–USST Joint Group for their valuable discussions on neuronal dynamics. This work was supported by the National Natural Science Foundation of China (Grant No. 121650165) and the Scientific Research Startup Fund of Jiangsu University (Grant No. 5501190017). M. Zheng acknowledges support from the Jiangsu Specially Appointed Professor Program.
\end{acknowledgements}


%
 \section*{Conflict of interest}
The authors declare that they have no conflict of interest.

\section{Appendixes}

\subsection{The parameter settings of the excitatory and inhibitory Gaussian kernels and their effects on the onset of seizure dynamics}

The spatial extent of the inhibitory field surrounding a focal discharge is approximately 
$3.0 $ $\mathrm{mm}$ \cite{prince1967control, schwartz2001vivo}. The electrophysiological data were obtained from patients with pharmacoresistant focal epilepsy who were additionally enrolled in studies employing microelectrode recordings of seizures over an area of $100 \times 100 $ $\mathrm{mm^2}$\cite{schevon2008microphysiology}. Thus, in the normalized spatial domain, the inhibitory spatial scale is given by $\sigma_I = 3 \mathrm{mm} / 100 \mathrm{mm} = 0.03$, consistent with seizure patient data reported in Ref.\cite{liou2020model}. 

\begin{figure}[htp]	
	\includegraphics[width=\linewidth]{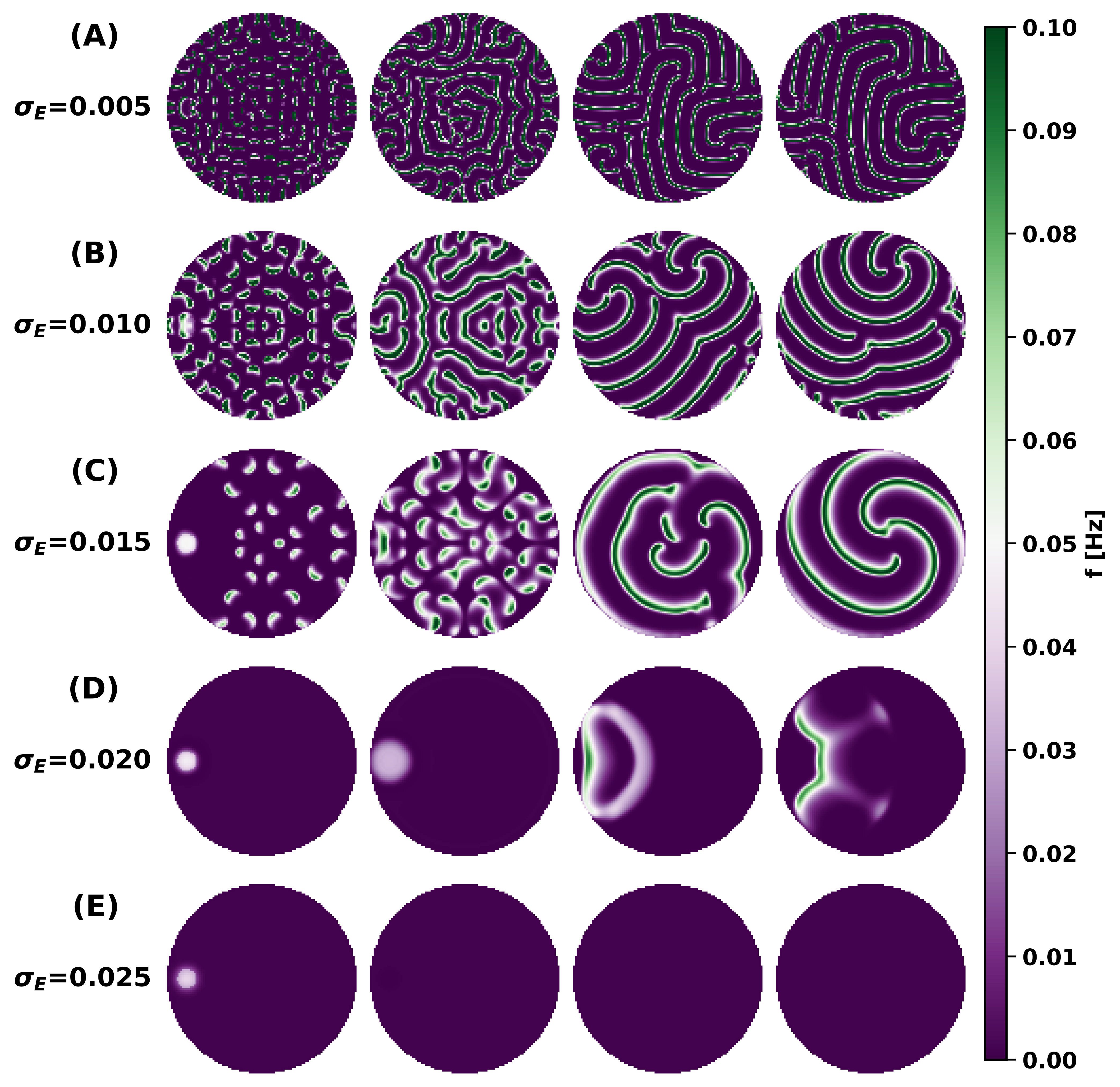}	  
	\caption{The spatiotemporal seizure firing patterns at different values of the standard deviation $\sigma_{E}$ when $\sigma_{I} = 0.03$ and $W_{con} = 1.0$. }
	\label{fig:sigmaE}
\end{figure}

To explore the effect of variance $\sigma_{E}^2$ on the arising seizure dynamics while keeping $\sigma_{I} = 0.03$,  Fig.\hyperref[fig:sigmaE]{\ref{fig:sigmaE}} illustrates various spatiotemporal seizure firing patterns at different values of the standard deviation $\sigma_{E}$. The neural network generates multi-arm spiral waves \cite{vasiev1997multiarmed} when $\sigma_{E} = 0.005$ and exhibits spiral waves when $\sigma_{E} = 0.010$ and  $\sigma_{E} = 0.015$. All of these firing patterns persist for extended periods without termination. More specifically, a transient excitatory perturbation initiates the ictal-tonic stage, followed by the ictal-clonic stage, and the post-ictal stage of seizure termination when $\sigma_{E} = 0.020$, as thoroughly investigated by Liou et al.\cite{liou2020model}. In contrast, the neural network fails to produce seizure dynamics when $\sigma_{E} = 0.025$. 

\subsection{Determining the ISI Threshold of firing rate through Histogram-Based ISI Distribution Analysis}
\begin{figure}[htp]	
	\includegraphics[width=\linewidth]{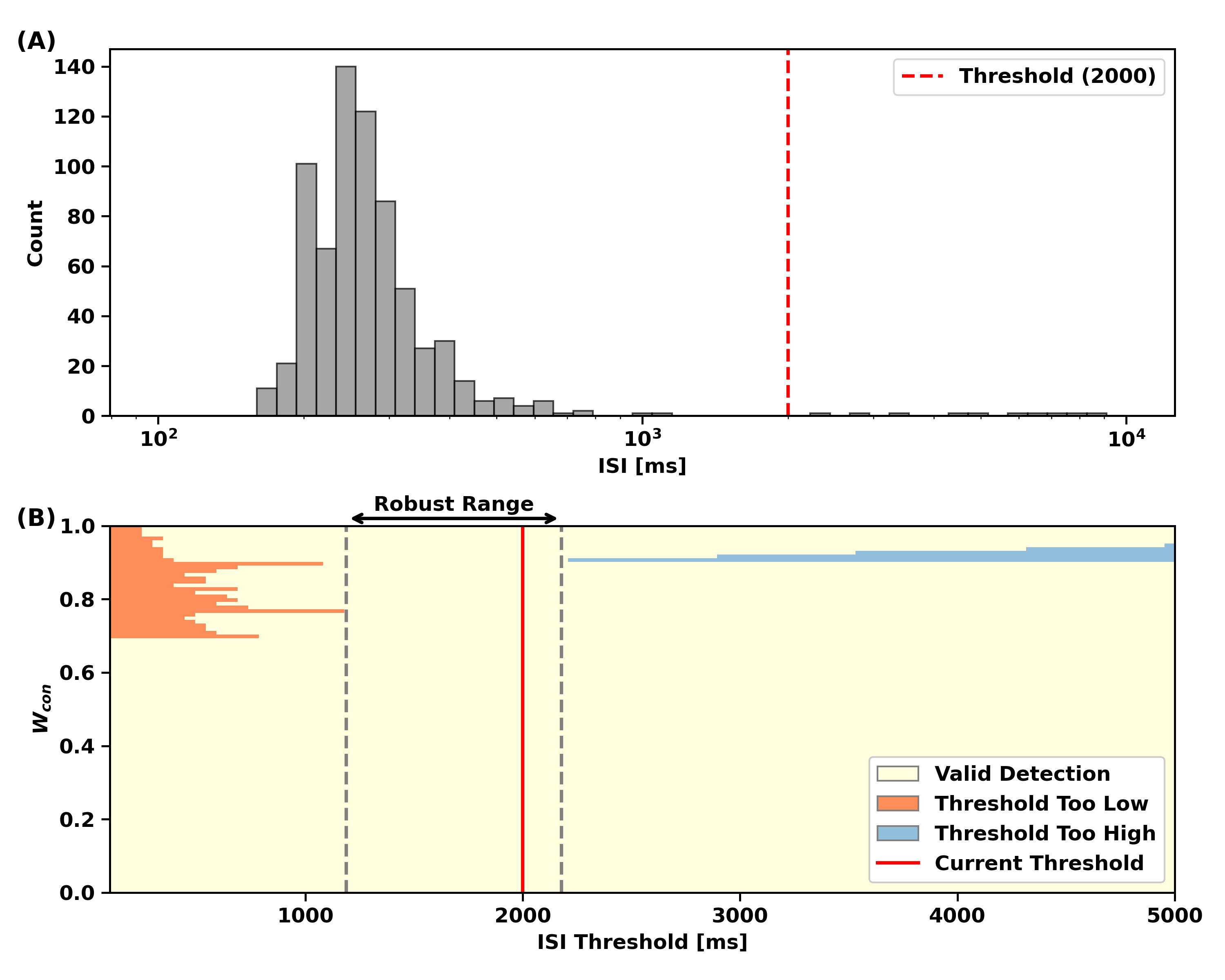}	  
	\caption{(A) Histogram of interspike intervals across the full range of $W_{con}$; (B) Choice of ISI threshold for seizure classification and transitions. }
	\label{fig:ISI}
\end{figure}

To determine an appropriate ISI threshold for distinguishing different seizure states, Fig.\hyperref[fig:ISI]{\ref{fig:ISI}A} shows the histogram of all interspike intervals collected from simulated seizures across the full range of $W_{con}\in[0,1.0]$. The ISI distribution exhibits a clear bimodal structure: the ictal–clonic phase is characterized by repetitive bursting with short intervals, forming a sharp peak around $\text{ISI}\approx 200$ ms, whereas the ictal–tonic phase corresponds to substantially longer intervals at $\text{ISI} >2000$ ms. A broad gap between these two separated components of the ISI distribution (approximately $1000$–$3000$ ms) contains almost no data points. Thus, an ISI threshold of $2000$ ms lies well within this gap and provides a statistically justified boundary for distinguishing ictal-tonic activity from ictal-clonic bursting.

To systematically assess whether the choice of ISI threshold affects seizure  classification and transitions, Fig.\hyperref[fig:ISI]{\ref{fig:ISI}B} presents a sensitivity analysis over the full range of $W_{con}\in{[0,1.0]}$. The yellow region in Fig.\hyperref[fig:ISI]{\ref{fig:ISI}B} indicates the parameter space in which the ISI threshold correctly identifies the presence or absence of the ictal–tonic phase. When the threshold is set too low ($\text{ISI}< 1000$ ms), slow ictal–clonic activity is incorrectly classified as ictal–tonic (orange region). In contrast, when the threshold is too high ($\text{ISI} > 3000$ ms), shorter ictal–tonic episodes are missed (light-blue region). By contrast, the robust range ($\text{ISI}\in[1200,2200]$ ms, between the two vertical dashed lines) reliably identifies seizure types and their transitions, independent of variations in $W_{con}$.

\subsection{Spiral wave formation and the measurement of spiral wave centers}\label{spirwave}

\begin{figure*}[htp]	
	\includegraphics[width=\linewidth]{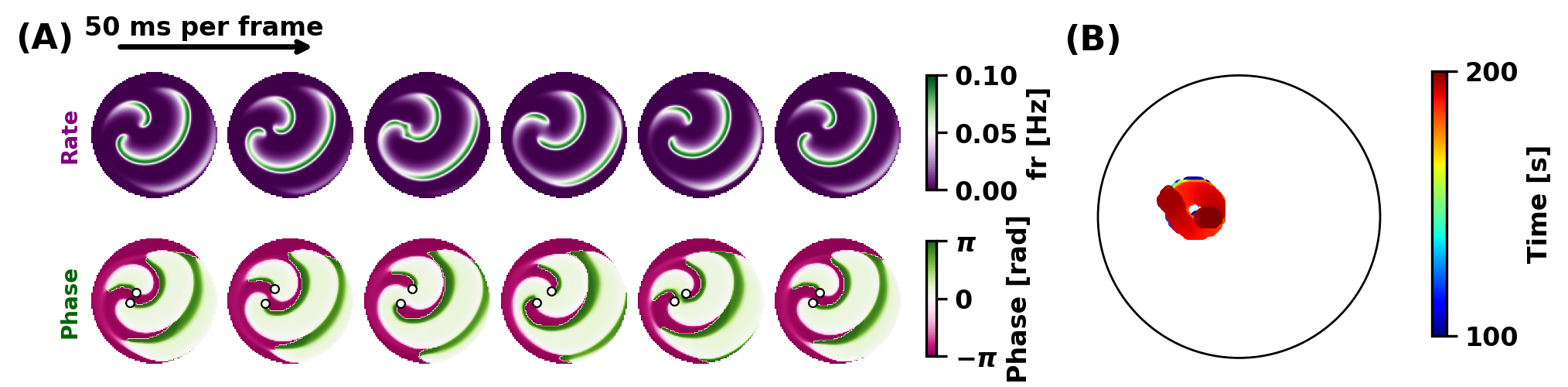}	  
	\caption{Spiral-wave formation in both the firing rate and the instantaneous phase (A), whose stability is quantified by identifying spiral-wave centers (B). }
	\label{fig:spiralwave}
\end{figure*}

Spiral-wave formation in the firing rate is quantified by identifying spiral-wave centers, corresponding to instantaneous phase singularities \cite{iyer2001experimentalist}. The instantaneous phase of the neuronal population at location $x$ and time $t$ is defined as $\psi(x,t)= \text{angle}(V(x,t)-V_m(x)+i(\phi(x,t)-\phi_m(x,t)))$ by the states of phase constructed in  the membrane potential and threshold plane parameter space.  Here, $V_m(x,t)$ and  $\phi_m(x,t)$ denote the median membrane potential and threshold, respectively, computed over the entire seizure episode at location $x$. Phase singularities are identified by detecting closed paths, $l$, encircling neighboring points for which $\oint_l \nabla \psi(x,t) \cdot dl = \pm 2\pi $.
Fig.\hyperref[fig:spiralwave]{\ref{fig:spiralwave}} shows spiral-wave dynamics in both the firing rate and the instantaneous phase (Fig.\hyperref[fig:spiralwave]{\ref{fig:spiralwave}A}). The corresponding spiral-wave centers accurately capture the stable rotation of the wave, as illustrated in Fig.\hyperref[fig:spiralwave]{\ref{fig:spiralwave}B}.

\subsection{Robustness of Seizure Dynamics to Propagation Delays}\label{taudelay}
\begin{figure}[htp]	
	\includegraphics[width=\linewidth]{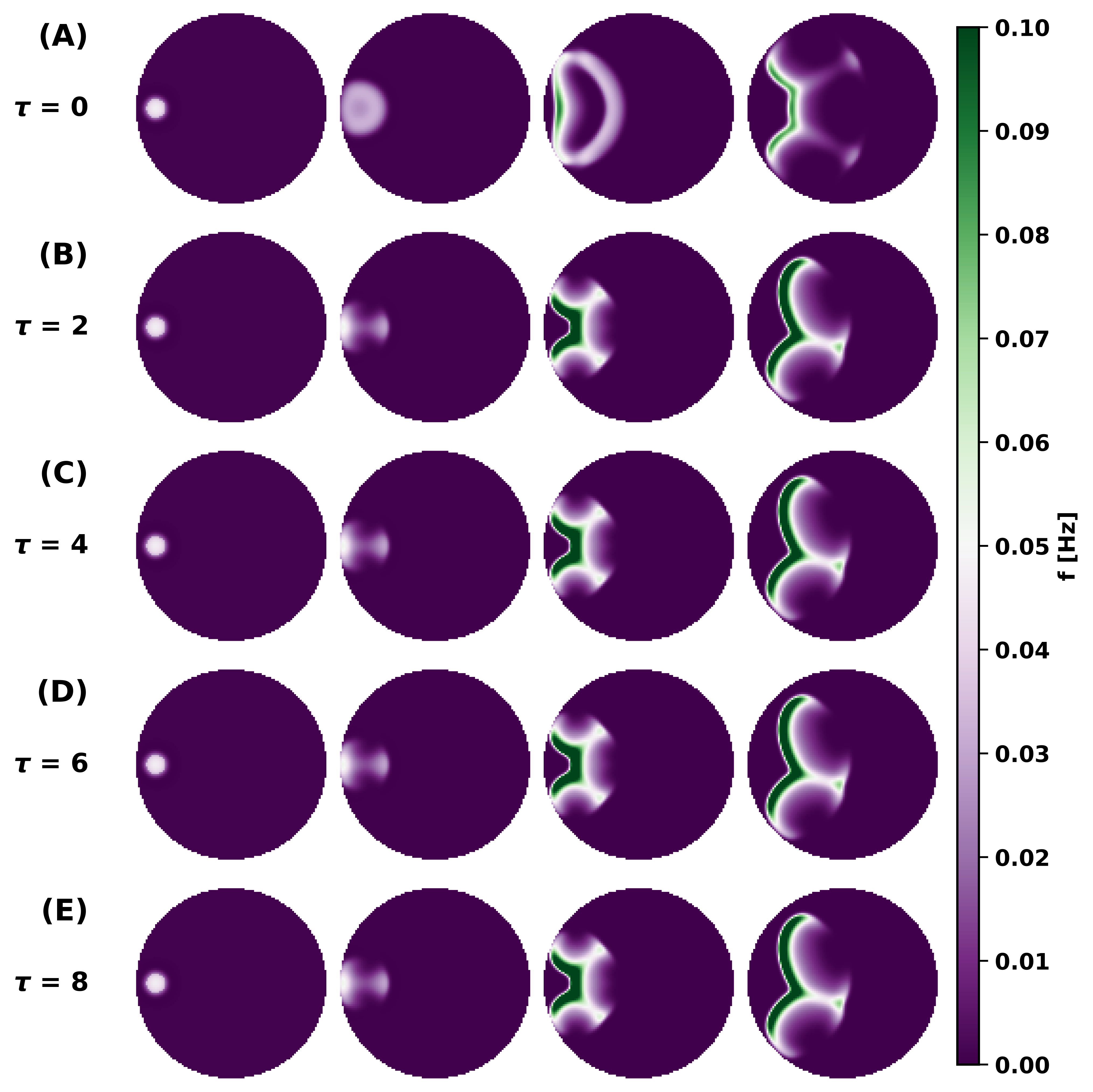}	  
	\caption{Spatiotemporal firing patterns across different propagation delays when $W_{con} = 1.0$. }
	\label{fig:delays}
\end{figure}

The propagation of membrane potentials between excitatory and inhibitory populations requires a finite delay $\tau$ which is regulated by inhibitory spiking activity. When $\tau = 0 $, excitatory and inhibitory neurons fire synchronously, corresponding to instantaneous responses between excitatory and inhibitory populations.   In contrast, $\tau > 0$ represents successive firing of excitatory and inhibitory neurons, with distinct firing frequencies that can be regulated by synaptic weights within the inhibitory population\cite{shan2025short,zhu2024chaos}. To represent this effect, we introduce a time delay by expressing the membrane potential of inhibitory neurons as $V_{ij}(t-\tau)$, where $V_{ij}$ denotes the membrane potential of the corresponding excitatory neurons. Fig.\hyperref[fig:delays]{\ref{fig:delays}} shows similar spatiotemporal firing patterns across increasing values of $\tau$, indicating that propagation delay makes only a minor contribution to the emergence of seizure dynamics in this work.

\subsection{Numerical Simulation}
The neural field model was implemented in MATLAB on a two-dimensional $N \times N$ nodes ($N=100$), masked by a circular domain of normalized radius 0.5, with activity outside the domain set to zero. Synaptic coupling was realized by convolving the firing rate with a Gaussian kernel using two-dimensional convolution with zero padding, followed by centering and cropping to the original grid size. The system was integrated using the exponential Euler scheme over 85 s with a time step $dt = 0.001$ s. The system was initialized at its equilibrium values: $V = -58$ mV, $\text{Cl}_{in} = 6$ mM,$\phi = -45$ and $g_K = 0$ nS.

\bibliographystyle{unsrt}
\bibliography{Controlseizure_revibib}   

%
%

\end{document}